\documentclass{article}
\usepackage{amssymb}
\usepackage{times}
\usepackage{amsmath}
\usepackage{amsfonts}
\usepackage{amsthm}
\usepackage{authblk}

\newtheorem{theorem}{Theorem}[section]

\newtheorem{corollary}[theorem]{Corollary}

\newtheorem{definition}[theorem]{Definition}

\newtheorem{lemma}[theorem]{Lemma}

\newtheorem{proposition}[theorem]{Proposition}

\newcommand{\la}{\langle}
\newcommand{\ra}{\rangle}

\newcommand{\cL}{{\mathcal L}}

\begin{document}

\title{Bounds on the Discrete Spectrum of Lattice Schr{\"o}dinger Operators}
\author{V.~Bach, W.~de Siqueira Pedra, S.N.~Lakaev}
\date{24-Sept-2017}
\maketitle

\begin{abstract}
We discuss the validity of the Weyl asymptotics -- in the sense of two-sided
bounds -- for the size of the discrete spectrum of (discrete) Schr{\"{o}}%
dinger operators on the $d$--dimensional, $d\geq 1$, cubic lattice $\mathbb{Z%
}^{d}$ at large couplings. We show that the Weyl asymptotics can be violated
in any spatial dimension $d\geq 1$ -- even if the semi-classical number of
bound states is finite. Furthermore, we prove for all dimensions $d\geq 1$
that, for potentials well-behaved at infinity and fulfilling suitable decay
conditions, the Weyl asymptotics always hold. These decay conditions are
mild in the case $d\geq 3$, while stronger for $d=1,2$. It is well-known
that the semi-classical number of bound states is -- up to a constant --
always an upper bound on the size of the discrete spectrum of Schr{\"{o}}%
dinger operators if $d\geq 3$. We show here how to construct general upper
bounds on the number of bound states of Schr{\"{o}}dinger operators on $%
\mathbb{Z}^{d}$ from semi-classical quantities in all space dimensions $%
d\geq 1$ and independently of the positivity-improving property of the free
Hamiltonian.\\[1.5ex]
\end{abstract}

\section{Introduction}

\label{sec-1}

Let $V\in L^{d/2}(\mathbb{R}^{d},\mathbb{R}_{0}^{+})$ be a non-negative
potential in the $d$--dimensional space with $d\geq 3$. From standard
results of spectral theory \cite{ReedSimonI-IV1980} it follows that the
negative spectrum $\sigma \lbrack -\Delta -\lambda V(x)]\cap \mathbb{R}^{-}$
of the corresponding self-adjoint Schr\"{o}dinger operator 
\begin{equation}
-\Delta _{\mathbb{R}^{d}}-\lambda V(x)  \label{eq-1.1}
\end{equation}%
on $L^{2}(\mathbb{R}^{d})$ is purely discrete, i.e., consists only of
isolated eigenvalues of finite multiplicity. Here, $\Delta _{\mathbb{R}%
^{d}}=\sum_{i=1}^{d}\partial _{x_{i}}^{2}$ is the Laplacian on $\mathbb{R}%
^{d}$ and $V$ acts as a multiplication operator, $[V{\varphi }](x)\doteq V(x)%
{\varphi }(x)$. By a well-known theorem -- first established by Weyl \cite%
{Weyl1911,Weyl1912b} for the case of Dirichlet Laplacian in a bounded domain
-- the number $N^{cont}[\lambda V]$ of negative eigenvalues of $-\Delta _{%
\mathbb{R}^{d}}-\lambda V$ (counting multiplicities) is asymptotically 
\begin{equation}
N^{cont}[\lambda V]\ \doteq \ \mathop{\mathrm{Tr}}\big\{\mathbf{1}[-\Delta _{%
\mathbb{R}^{d}}-\lambda V<0]\big\}\ \sim \ N_{sc}^{cont}[\lambda V]
\label{eq-1.2}
\end{equation}%
as $\lambda \rightarrow \infty $. The right side of \eqref{eq-1.2} is the
volume 
\begin{equation}
N_{sc}^{cont}[V]\ \doteq \ \int \mathbf{1}[p^{2}-V(x)<0]\,\frac{\mathrm{d}%
^{d}x\,\mathrm{d}^{d}p}{(2\pi )^{d}}  \label{eq-1.3}
\end{equation}%
these bound states occupy in phase space $\mathbb{R}^{d}\times (\mathbb{R}%
^{\ast })^{d}=\mathbb{R}^{2d}$ according to semi-classical analysis. This
so-called Weyl asymptotics \eqref{eq-1.2} is complemented by the celebrated 
\textit{non-asymptotic} bound of Rozenblum \cite{Rozenblum1972}, Lieb \cite%
{Lieb1976b}, and Cwikel \cite{Cwikel1977} on the number $N^{cont}[V]$ of
bound states of $-\Delta _{\mathbb{R}^{d}}-V$ of the form 
\begin{equation}
N^{cont}[V]\ \leq \ C_{\mathrm{CLR}}(d)\,N_{sc}^{cont}[V]  \label{eq-1.4}
\end{equation}%
for some $C_{\mathrm{CLR}}(d)\geq 1$. Lieb \cite[Eq.~(4.5)]{Lieb1980a} has
shown that the optimal choice for $C_{\mathrm{CLR}}(3)$ is smaller than $6.9$%
. Note that 
\begin{equation}
N_{sc}^{cont}[V]\ =\ \frac{|S^{d-1}|}{d\,(2\pi )^{d}}\int V^{d/2}(x)\,%
\mathrm{d}^{d}x,  \label{eq-1.5}
\end{equation}%
where $|S^{d-1}|$ is the volume of the $(d-1)$--dimensional sphere.

In the present paper, we replace the Euclidean $d$--dimensional space $%
\mathbb{R}^{d}$ by the $d$--dimensional hypercubic lattice $\Gamma =\mathbb{Z%
}^{d}$ and study the discrete analogues of the Weyl asymptotics %
\eqref{eq-1.2} and the Cwikel-Lieb-Rozenblum (CLR) bound \eqref{eq-1.4}. For
a given \textit{potential} $V\in \ell ^{\infty }(\Gamma ,\mathbb{R}_{0}^{+})$%
, the discrete Schr\"{o}dinger operator corresponding to \eqref{eq-1.1} is 
\begin{equation}
-\Delta _{\Gamma }-\lambda V(x),  \label{eq-1.6}
\end{equation}%
where $V$ acts again as a multiplication operator and $\Delta _{\Gamma }$ is
the discrete Laplacian defined by 
\begin{equation}
\lbrack \Delta _{\Gamma }{\varphi }](x)\ =\ \sum_{|v|=1}\big\{{\varphi }(x)-{%
\varphi }(x+v)\big\}.  \label{eq-1.7}
\end{equation}%
More generally, we assume to be given a Morse function $\mathfrak{e}\in
C^{2}(\Gamma ^{\ast },\mathbb{R})$ on the $d$--dimensional torus (Brillouin
zone) $\Gamma ^{\ast }=\big(\mathbb{R}/2\pi \mathbb{Z}\big)^{d}=[-\pi ,\pi
)^{d}$, the dual group of $\Gamma $. Given such a function $\mathfrak{e}$,
we consider the self-adjoint operator 
\begin{equation}
H(\mathfrak{e},V)\ \doteq \ h(\mathfrak{e})-V(x),  \label{eq-1.8}
\end{equation}%
on $\ell ^{2}(\Gamma )$, where $h(\mathfrak{e})\in \mathcal{B}[\ell
^{2}(\Gamma )]$ is the hopping matrix (convolution operator) corresponding
to the dispersion relation $\mathfrak{e}$, i.e., 
\begin{equation}
\big[\mathcal{F}^{\ast }\big(h(\mathfrak{e}){\varphi }\big)\big](p)\ =\ 
\mathfrak{e}(p)\,[\mathcal{F}^{\ast }({\varphi })](p)\;,  \label{eq-1.9}
\end{equation}%
for all ${\varphi }\in L^{2}(\Gamma ^{\ast })$ and all $p\in \Gamma ^{\ast }$%
. Here, 
\begin{equation}
\mathcal{F}^{\ast }:\ \ell ^{2}(\Gamma )\rightarrow L^{2}(\Gamma ^{\ast
}),\quad \lbrack \mathcal{F}^{\ast }({\varphi })](p)\ \doteq \ \sum_{x\in
\Gamma }\mathrm{e}^{-i\langle p,x\rangle }{\varphi }(x)  \label{eq-1.10}
\end{equation}%
is the usual discrete Fourier transformation with inverse 
\begin{equation}
\mathcal{F}:\ L^{2}(\Gamma ^{\ast })\rightarrow \ell ^{2}(\Gamma ),\quad
\lbrack \mathcal{F}(\psi )](x)\ \doteq \ \int_{\Gamma ^{\ast }}\mathrm{e}%
^{i\langle p,x\rangle }\psi (p)\,\mathrm{d}\mu ^{\ast }(p)\,,
\label{eq-1.11}
\end{equation}%
where $\mu ^{\ast }$ is the (normalized) Haar measure on the torus, $\mathrm{%
d}\mu ^{\ast }(p)=\frac{\mathrm{d}^{d}p}{(2\pi )^{d}}$. Put differently, $h(%
\mathfrak{e})=\mathcal{F}\mathfrak{e}\mathcal{F}^{\ast }$ is the Fourier
multiplier corresponding to $\mathfrak{e}$. We assume w.l.o.g.\ that the
minimum of $\mathfrak{e}$ is $0$, so 
\begin{equation}
\mathfrak{e}(\Gamma ^{\ast })\ =\ [0,{\mathfrak{e}_{\mathrm{max}}}(\mathfrak{%
e})],  \label{eq-1.12}
\end{equation}%
and we call a Morse function $\mathfrak{e}\in C^{2}(\Gamma ^{\ast },\mathbb{R%
})$ obeying \eqref{eq-1.12} an \textit{admissible dispersion relation}. Note
that $-\Delta _{\Gamma }=h(\mathfrak{e}_{\mathrm{Lapl}})$, with 
\begin{equation}
\mathfrak{e}_{\mathrm{Lapl}}(p)\ \doteq \ \sum_{i=1}^{d}\big(1-\cos (p_{i})%
\big)\;,\quad \mathfrak{e}_{\mathrm{Lapl}}(\Gamma ^{\ast })\ =\ [0,2d],
\label{eq-1.13}
\end{equation}%
being admissible. We require that $V$ decays at infinity, 
\begin{equation}
V\ \in \ \ell _{0}^{\infty }(\Gamma ,\mathbb{R}_{0}^{+})\ \doteq \ \Big\{%
V:\Gamma \rightarrow \mathbb{R}_{0}^{+}\;|\ \lim_{|x|\rightarrow \infty
}V(x)=0\Big\},  \label{eq-1.14}
\end{equation}%
or sometimes even that $V$ has bounded support. Note that $V\in \ell
_{0}^{\infty }(\Gamma ,\mathbb{R}_{0}^{+})$ is compact as a multiplication
operator on $\ell ^{2}(\Gamma )$ and by a(nother) theorem of Weyl, 
\begin{equation}
\sigma _{\mathrm{ess}}[H(\mathfrak{e},V)]\ =\ \sigma _{\mathrm{ess}}[H(%
\mathfrak{e},0)]\ =\ [0,{\mathfrak{e}_{\mathrm{max}}}],  \label{eq-1.15}
\end{equation}%
where {$\mathfrak{e}$}${_{\mathrm{max}}}\equiv \,${$\mathfrak{e}$}${_{%
\mathrm{max}}}(\mathfrak{e})$. From the positivity of $V$ and the min-max
principle we further obtain that all isolated eigenvalues of finite
multiplicity lie below $0$, 
\begin{equation}
\sigma _{\mathrm{disc}}[H(\mathfrak{e},V)]\ \subseteq \ \mathbb{R}^{-}\
\doteq \ (-\infty ,0).  \label{eq-1.16}
\end{equation}%
We note in passing that -- different to Schr\"{o}\-din\-ger operators on $%
\mathbb{R}^{d}$ -- discrete Schr\"{o}\-din\-ger operators possibly have
positive eigenvalues when changing the sign of the potential. Counting the
number of positive eigenvalues, however, can be traced back to the case
treated here by replacing $\mathfrak{e}(p)$ by {$\mathfrak{e}$}${_{\mathrm{%
max}}}-\mathfrak{e}(p)$.

Our goal in this paper is to give -- in all dimensions -- both asymptotic
and non-asymptotic bounds on the number 
\begin{equation}
N[\mathfrak{e},V]\ \doteq \ \mathop{\mathrm{Tr}}\big\{\mathbf{1}[H(\mathfrak{%
e},V)<0]\big\}  \label{eq-1.17}
\end{equation}%
of negative eigenvalues of $H(\mathfrak{e},V)$. Criteria for $N[\mathfrak{e}%
,V]$ to be finite or to be infinite were given in \cite%
{AbdullaevIkromov2007, Karachalios2008, RozenblumSolomyak2008}. The main
focus lies on the physically most relevant case $d \geq 3$. For $d=1$ the
situation is well-understood \cite{DamanikHundertmarkKillipSimon2003,
DamanikKillipSimon2005}, and even asymptotics for the accumulation of
eigenvalues near zero are known \cite{DamanikTeschl2007}. The case $d =2$ is
particularly difficult, see \cite{RozenblumSolomyak2013} for the best
results currently known.

In the present paper, we aim at bounds on $N[\mathfrak{e},V]$ in terms of
the corresponding semi-classical quantity 
\begin{eqnarray}
N_{sc}[\mathfrak{e},V] &\doteq &\sum_{x\in \Gamma }\int_{\Gamma ^{\ast }}%
\mathbf{1}[\mathfrak{e}(p)-V(x)<0]\text{ }\mathrm{d}\mu ^{\ast }(p) \\
&=&\int_{\Gamma ^{\ast }}\mathcal{L}_{V}[\mathfrak{e}(p)]\text{ }\mathrm{d}%
\mu ^{\ast }(p),
\end{eqnarray}%
where the sizes $\mathcal{L}_{V}[\alpha ]\in \mathbb{N}_{0}$ of the level
sets of $V$ are defined by 
\begin{equation}
\mathcal{L}_{V}[\alpha ]\ \doteq \ \#\big\{x\in \Gamma \;\big|\ V(x)\geq
\alpha \big\}  \label{eq-1.19}
\end{equation}%
for $\alpha >0$. Note that $\mathcal{L}_{V}[\alpha ]$ is independent of the
localization properties of $V$. This lets us introduce the notion of
rearrangements of $V$. Given $V,{\widetilde{V}}\in \ell _{0}^{\infty
}(\Gamma ,\mathbb{R}_{0}^{+})$, we say that 
\begin{equation}
\text{${\widetilde{V}}$ is a \textit{rearrangement} of $V$, whenever }%
\forall \,\alpha >0:\ \mathcal{L}_{{\widetilde{V}}}[\alpha ]=\mathcal{L}%
_{V}[\alpha ].  \label{eq-1.20}
\end{equation}%
In other words, the supports of$\mathrm{\;}{\widetilde{V}}$ and ${\widetilde{%
V}}$ have the same cardinality, and ${\widetilde{V}|}_{\mathrm{supp\;}{%
\widetilde{V}}}=V\circ J$ for some bijection $J:\mathrm{supp\;}{\widetilde{V}%
}\rightarrow \mathrm{supp\;}{V}$. Obviously, being rearrangements of each
other defines an equivalence relation on $\ell _{0}^{\infty }(\Gamma ,%
\mathbb{R}_{0}^{+})$. The importance of the growth of the sizes $\mathcal{L}%
_{V}[\alpha ]$ of the level sets of $V$, as $\alpha \searrow 0$, is also
realized in \cite{RozenblumSolomyak2008, RozenblumSolomyak2009}.

We emphasize that in most other studies of $N[\mathfrak{e},V]$ and notably
in \cite{RozenblumSolomyak2008, RozenblumSolomyak2009,
RozenblumSolomyak2010, RozenblumSolomyak2011}, the generator of the kinetic
energy is assumed to be Markovian. By constrast, we use CLR\ bounds recently
derived in \cite{Frank2014} that do not require such an assumption and the
only essential property of the dispersion $\mathfrak{e}$ we need in our
proofs is its Morse property.

\subsection{Non-Asymptotic Semi-classical Bounds}

\label{subsec-1.1}

We first formulate our non-asymptotic results which correspond to the CLR
bound \eqref{eq-1.4} in the continuum case.

\begin{theorem}[Non-asymptotic upper bound, $d\geq 3$]
\label{thm-1.1} Let $d\geq 3$ and $\mathfrak{e}$ an admissible dispersion.
Then there exists a constant $C_{\ref{thm-1.1}}(d,\mathfrak{e})\in \lbrack
1,\infty )$ such that 
\begin{equation}
N[\mathfrak{e},V]\ \leq \ C_{\ref{thm-1.1}}(d,\mathfrak{e})\,N_{sc}[%
\mathfrak{e},V]\ <\ \infty   \label{eq-1.22}
\end{equation}%
for all $V\in \ell ^{d/2}(\Gamma ,\mathbb{R}_{0}^{+})$.
\end{theorem}

If $d<3$, the following weighted version of the non-asymptotic bound on $N[%
\mathfrak{e},V]$ still holds:

\begin{theorem}[Non-asymptotic upper bound, $d<3$]
\label{thm-1.2} Let $d\in \{1,2\}$ and assume that $\mathfrak{e}\in
C^{3}(\Gamma ^{\ast },\mathbb{R}_{0}^{+})$. Then there is a constant $C_{\ref%
{thm-1.2}}(d,\mathfrak{e})<\infty $ such that, for any potential $V\in \ell
_{0}^{\infty }(\Gamma ,\mathbb{R}_{0}^{+})$, 
\begin{equation}
N[\mathfrak{e},V]\ \leq \ C_{\ref{thm-1.2}}(d,\mathfrak{e})\Big(1+N_{sc}\big[%
\mathfrak{e},\,|x|^{d+5}V\big]\Big).  \label{eq-1.23}
\end{equation}
\end{theorem}

Our results show that the right quantity to compare the number of
eigenvalues to is the phase space volume $N_{sc}[\mathfrak{e},V]$ of the set 
$\{(p,x)\;|\;\mathfrak{e}(p)-V(x)<0\}$ and not (the $\frac{d}{2}^{th}$ power
of) the $\ell ^{d/2}$--norm of $V$. In the case of Schr\"{o}dinger operators
on $\mathbb{R}^{d}$, these quantities agree up to a multiplicative constant.
See \eqref{eq-1.5}. While it is possible to bound $N_{sc}[\mathfrak{e},V]$
and hence also $N[\mathfrak{e},V]$ by a multiple of $|V|_{d/2}^{d/2}=%
\sum_{x}V^{d/2}(x)$, this bound grossly overestimates the number of
eigenvalues in the limit of large couplings. For example, if $\Lambda
\subset \Gamma $ is a finite subset then 
\begin{equation}
N_{sc}[\mathfrak{e},\lambda \,\mathbf{1}_{\Lambda }]\ =\ |\Lambda |\ \ll \
\lambda ^{d/2}\,|\Lambda |\ =\ |\lambda \,\mathbf{1}_{\Lambda }|_{d/2}^{d/2}
\label{eq-1.24}
\end{equation}%
for sufficiently large $\lambda >0$.

In Sect.~\ref{subsec-3.2} we prove the optimality of Theorem~\ref{thm-1.1}
with respect to the class $\ell^{d/2}(\Gamma ,\mathbb{R}_{0}^{+})\ni V$.

\begin{theorem}[Optimality of $\ell^{d/2}(\Gamma ,\mathbb{R}_{0}^{+})\ni V$
in Thm.~\protect\ref{thm-1.1}]
\label{thm-1.3} Let $d\geq 1$ and $\mathfrak{e}$ be an admissible dispersion
for which $|h(\mathfrak{e})_{0,x}|\leq \mathrm{const\,}\langle x\rangle
^{-2(d+1)}$ for some $\mathrm{const}<\infty $. Then, for any ${\varepsilon }%
>0$, there exists a potential $V_{\varepsilon }\in \ell^{(d/2)+{\varepsilon }%
}(\Gamma ,\mathbb{R}_{0}^{+})\setminus \ell^{d/2}(\Gamma ,\mathbb{R}%
_{0}^{+}) $ such that $N[\mathfrak{e},V_{\varepsilon }]=N_{sc}[\mathfrak{e}%
,V_{\varepsilon }]=\infty $.

Here, $h(\mathfrak{e})_{x,y}\doteq \big|\big\la\delta _{x}\big|\,h({%
\mathfrak{e}})\delta _{y}\big\ra\big|$ are the matrix elements w.t.r. to the
canonical basis of $\ell^{2}(\Gamma )$ of the hopping matrix $h({\mathfrak{e}%
})$ of the dispersion relation $\mathfrak{e}$, and $\langle x\rangle \doteq
1+|x|$.
\end{theorem}

This does not, however, imply that $N[\mathfrak{e},V]=\infty $ whenever $%
N_{sc}[\mathfrak{e},V]=\infty $. For instance, if $V(x)=\langle x\rangle
^{-2}(\log \langle x\rangle )^{-\eta }$ for some $\eta \in (0,2/d)$ then $N[%
\mathfrak{e}_{\mathrm{Lapl}},V]<\infty $ but $N_{sc}[\mathfrak{e}_{\mathrm{%
Lapl}},V]=\infty $. See the example in \cite[Section~5.2]%
{RozenblumSolomyak2009}.

We complement the non-asymptotic upper bounds by corresponding lower bounds:

\begin{theorem}[Non-Asymptotic Lower Bound]
\label{thm-1.4} Let $d\geq 1$ and $\mathfrak{e}$ be an admissible
dispersion. Then, for any potential $V\in \ell _{0}^{\infty }(\Gamma ,%
\mathbb{R}_{0}^{+})$ and all $c>{\mathfrak{e}_{\mathrm{max}}}$, 
\begin{equation}
N[\mathfrak{e},V]\ \geq \ \mathcal{L}_{V}[c]\ =\ \#\big\{x\in \Gamma \,|\
V(x)\geq c\big\}.  \label{eq-1.26}
\end{equation}
\end{theorem}

From Theorems~\ref{thm-1.1} and \ref{thm-1.4} emerges the interesting
question, whether $N_{sc}[\mathfrak{e},V]$ or $\mathcal{L}_{V}[${$\mathfrak{e%
}$}${_{\mathrm{max}}}]$ (or both) are saturated in certain limits. It turns
out that, for sparse potentials $V$, the number $N[\mathfrak{e},V]$ bound
states is correctly described by $\mathcal{L}_{V}[\eta (\mathfrak{e})]$,
where $0\leq \eta (\mathfrak{e})<\mathfrak{e}_{\mathrm{max}}$ is defined by 
\begin{equation}
\frac{1}{\eta (\mathfrak{e})}\ \doteq \ \int_{\Gamma ^{\ast }}\frac{\mathrm{d%
}\mu ^{\ast }(p)}{\mathfrak{e}(p)}\,,  \label{eq-1.25}
\end{equation}%
for $d\geq 3$, and $\eta (\mathfrak{e})\doteq 0$ for $d\in \{1,2\}$ -- see,
for instance, Lemma~\ref{lem-3.6.b} and the proof of Corollary~\ref%
{cor-4.5,1}. Observe that, as $\eta (\mathfrak{e})\leq {\mathfrak{e}_{%
\mathrm{max}}}$, $\mathcal{L}_{V}[${$\mathfrak{e}$}${_{\mathrm{max}}}]$ $%
\leq \mathcal{L}_{V}[\eta (\mathfrak{e})].$ Since for any $\delta >0$ there
is a potential $V\in \ell _{0}^{\infty }(\Gamma ,\mathbb{R}_{0}^{+})$ for
which $\mathcal{L}_{V}[\eta (\mathfrak{e})]/\mathcal{L}_{V}[${$\mathfrak{e}$}%
${_{\mathrm{max}}}]<1+\delta $, the following theorem implies the optimality
of the lower bound in Theorem~\ref{thm-1.4} with respect to rearrangements.

\begin{theorem}[Optimality of Thm.~\protect\ref{thm-1.4} under rearrangements%
]
\label{thm-1.5} Let $d\geq 3$, $\mathfrak{e}$ be an admissible dispersion.
Given ${\varepsilon }\in (0,1)$ and a potential $V\in \ell _{0}^{\infty
}(\Gamma ,\mathbb{R}_{0}^{+})$, there exists a rearrangement ${\widetilde{V}}%
\in \ell _{0}^{\infty }(\Gamma ,\mathbb{R}_{0}^{+})$ of $V$ such that 
\begin{equation}
N[\mathfrak{e},{\widetilde{V}}]\ \leq \ \mathcal{L}_{{\widetilde{V}}}[(1-{%
\varepsilon })\eta (\mathfrak{e})]\ =\ \#\big\{x\in \Gamma \,\big|\ V(x)\geq
(1-{\varepsilon })\eta (\mathfrak{e})\big\}.  \label{eq-1.27}
\end{equation}
\end{theorem}

In general, the semi-classical number of bound states $\,N_{sc}[\mathfrak{e}%
,\lambda V]$ is not a lower bound on $N[\mathfrak{e},\lambda V]$ -- not even
up to prefactors. This is illustrated by the following theorem.

\begin{theorem}
\label{thm-1.9} Let $d\geq 3$ and $\mathfrak{e}$ be an admissible
dispersion. Then there exists a potential $V\notin \bigcup\limits_{p\geq
1}\ell^{p}(\Gamma )$ with $N[\mathfrak{e},V]=0$.
\end{theorem}

\subsection{(Weyl-)Asymptotic Semi-classical Bounds}

\label{subsec-1.2}

The Weyl asymptotics \eqref{eq-1.2} states that, for all fixed potentials $%
V\in L^{d/2}(\mathbb{R}^{d})$, 
\begin{equation}
\lim_{\lambda \rightarrow \infty }\frac{N^{cont}[\lambda V]}{%
N_{sc}^{cont}[\lambda V]}\ =\ 1,  \label{eq-1.28}
\end{equation}%
and that $N_{sc}^{cont}[\lambda V]=\lambda^{d/2}\,N_{sc}^{cont}[V]$. For
discrete Schr\"{o}dinger operators, only weaker statements hold true, as is
illustrated by the following lemma.

\begin{lemma}
\label{lem-1.6} Assume $d\geq 3$ and $V\in \ell ^{d/2}(\Gamma ,\mathbb{R}%
_{0}^{+})$. Then 
\begin{equation}
\lim_{\lambda \rightarrow \infty }\big\{\lambda ^{-d/2}\,N[\mathfrak{e}%
,\lambda V]\big\}\ =\ \lim_{\lambda \rightarrow \infty }\big\{\lambda
^{-d/2}\,N_{sc}[\mathfrak{e},\lambda V]\big\}\ =\ 0.  \label{eq-1.28,1}
\end{equation}
\end{lemma}

For a precise formulation of our asymptotic bounds, we introduce the numbers 
\begin{eqnarray}
g_{+}(V) &\doteq &\sup_{r>0}\limsup_{\ell \rightarrow \infty }\frac{2}{d\,r}%
\Big(\ln \mathcal{L}_{V}\big[\mathrm{e}^{-\ell -r}\big]-\ln \mathcal{L}_{V}%
\big[\mathrm{e}^{-\ell }\big]\Big),  \label{eq-1.30} \\[1ex]
g_{-}(V) &\doteq &\inf_{r>0}\liminf_{\ell \rightarrow \infty }\frac{2}{d\,r}%
\Big(\ln \mathcal{L}_{V}\big[\mathrm{e}^{-\ell -r}\big]-\ln \mathcal{L}_{V}%
\big[\mathrm{e}^{-\ell }\big]\Big),
\end{eqnarray}%
built from the level sets of $V$. While the significance of $g_{-}(V)$ is
made clear in Section~\ref{subsec-4.1}, $g_{+}(V)$ directly enters the
following theorem.

\begin{theorem}[Asymptotic bounds, $d\geq 3$]
\label{thm-1.7} Assume $d\geq 3$ and $V\in \ell ^{d/2}(\Gamma ,\mathbb{R}%
_{0}^{+})$. Then there are constants $0<C_{\ref{thm-1.7}s}(d,\mathfrak{e}%
)\leq C_{\ref{thm-1.7}g}(d,\mathfrak{e})<\infty $ such that 
\begin{eqnarray}
\big(1-g_{+}(V)\big)\,C_{\ref{thm-1.7}s}(d,\mathfrak{e}) &\leq
&\liminf_{\lambda \rightarrow \infty }\bigg\{\frac{N[\mathfrak{e},\lambda V]%
}{N_{sc}[\mathfrak{e},\lambda V]}\bigg\}  \label{eq-1.31} \\
&\leq &\limsup_{\lambda \rightarrow \infty }\bigg\{\frac{N[\mathfrak{e}%
,\lambda V]}{N_{sc}[\mathfrak{e},\lambda V]}\bigg\}\ \;\leq \ \;C_{\ref%
{thm-1.7}g}(d,\mathfrak{e}).\hspace{10mm}  \notag
\end{eqnarray}
\end{theorem}

A somewhat weaker form of Theorem~\ref{thm-1.7} still holds in case $d<3$.

\begin{theorem}[Asymptotic Bounds, \ $d<3$]
\label{thm-1.8} Assume that $d\in \{1,2\}$ and $V\in \ell ^{d/2}(\Gamma ,%
\mathbb{R}_{0}^{+})$. Then there are constants $0<C_{\ref{thm-1.8}s}(d,%
\mathfrak{e})\leq C_{\ref{thm-1.8}g}(d,\mathfrak{e})<\infty $ such that 
\begin{eqnarray}
\big(1-g_{+}(V)\big)\,C_{\ref{thm-1.8}s}(d,\mathfrak{e}) &\leq
&\liminf_{\lambda \rightarrow \infty }\bigg\{\frac{N[\mathfrak{e},\lambda V]%
}{N_{sc}[\mathfrak{e},\lambda V]}\bigg\},  \label{eq-1.33} \\
\limsup_{\lambda \rightarrow \infty }\bigg\{\frac{N[\mathfrak{e},\lambda V]}{%
1+N_{sc}[\mathfrak{e},\lambda |x|^{d+5}V]}\bigg\} &\leq &C_{\ref{thm-1.8}%
g}(d,\mathfrak{e}).\hspace{10mm}  \notag
\end{eqnarray}
\end{theorem}

We remark that if $V$ is rapidly decaying then, typically, $g_{+}(V)=0$. For
instance, if 
\begin{equation}
c_{1}\,\mathrm{e}^{-\alpha _{1}|x|}\ \leq \ V(x)\ \leq \ c_{2}\,\mathrm{e}%
^{-\alpha _{2}|x|}\;,  \label{eq-1.35}
\end{equation}%
for some constants $c_{1},\alpha _{1},\alpha _{2}>0$, $c_{2}<\infty $, and
all $x\in \Gamma $, then $g_{+}(V)=0$. Moreover, by the bounds proven here,
in this case the usual Weyl semi-classical asymptotics hold true in all
dimensions $d\geq 1$ and for all admissible dispersion relations, in the
sense that 
\begin{equation}
\lim_{\lambda \rightarrow \infty }\bigg\{\frac{N[\mathfrak{e},\lambda V]}{%
N_{sc}[\mathfrak{e},\lambda V]}\bigg\}\ =\lim_{\lambda \rightarrow \infty }%
\bigg\{\frac{N[\mathfrak{e},\lambda V]}{1+N_{sc}[\mathfrak{e},\lambda
|x|^{^{\alpha _{d}}}V]}\bigg\}\ =\ 1.  \label{eq-1.36}
\end{equation}%
We further remark that if $V$ behaves at infinity like an inverse power of $%
|x|$, i.e., if the limit 
\begin{equation}
\lim_{|x|\rightarrow \infty }\bigg\{\frac{-\log [V(x)]}{\log |x|}\bigg\}\ =\
\beta   \label{eq-1.37}
\end{equation}%
exists, then $g_{+}(V)=g_{-}(V)=2\beta /d$. In particular, in this case $%
g_{+}(V)<1$ implies $V\in \ell ^{d/2}(\Gamma ,\mathbb{R}_{0}^{+})$, and $%
g_{-}(V)>1$ implies $V\notin \ell ^{d/2}(\Gamma ,\mathbb{R}_{0}^{+})$.

In contrast to the continuum case, the boundedness of $V$ in $\ell^{d/2}$
alone does not suffice to ensure the semi-classical asymptotic behavior of $%
N[\mathfrak{e},\lambda V]$, but details of the behavior of $V$ at infinity
enter, too, as is illustrated by the following theorem.

\begin{theorem}
\label{thm-liminf-limsup} Let $d\geq 3$ and $\mathfrak{e}$ be an admissible
dispersion. Then there exists a potential $V$ with $N_{sc}[\mathfrak{e}%
,\lambda V]<\infty $ for all $\lambda >0$ for which%
\begin{equation*}
\liminf_{\lambda \rightarrow \infty }\frac{N_{sc}[\mathfrak{e},\lambda V]}{N[%
\mathfrak{e},\lambda V]}<\infty \;\ \ \mathrm{and}\text{\ \ \ }%
\limsup_{\lambda \rightarrow \infty }\frac{N_{sc}[\mathfrak{e},\lambda V]}{N[%
\mathfrak{e},\lambda V]}=\infty {.}
\end{equation*}
\end{theorem}

In fact, potentials on the lattice can be so peculiar that their eigenvalue
asymptotics assumes any prescribed behavior in the sense of the following
theorem.

\begin{theorem}
\label{thm-1.10} Let $d\geq 3$ and $\mathfrak{e}$ be any admissible
dispersion. Let further $F:[1,\infty )\rightarrow \mathbb{N}$ be an
arbitrary monotonically increasing, positive, integer-valued,
right-continuous function. Then, for any $\varepsilon \in (0,1/2)$, there
exists a potential $V_{F,\varepsilon }\in \ell _{0}^{\infty }(\Gamma )$ such
that 
\begin{equation}
\forall \, {\lambda \geq 2}: \quad F\big((1-\varepsilon )\lambda \big)\ \leq
\ N\big[\mathfrak{e},\lambda V_{F,\varepsilon }\big]\ \leq \ F\big(%
(1+\varepsilon )\lambda \big).  \label{eq-1.37,5}
\end{equation}
\end{theorem}

Results similar to Theorems~\ref{thm-1.7}, \ref{thm-liminf-limsup}, and \ref%
{thm-1.10} have been obtained in \cite{RozenblumSolomyak2008,
RozenblumSolomyak2009}, where the property $g_{+}(V) < 1$ has been
characterized by $V \in \ell_{q,w}(\mathbb{Z}^d)$ belonging to a weak $%
\ell_{q}$-space, for some $q > d/2$. The latter ensures that $\mathcal{L}%
_{V}[\alpha] \leq C \cdot \alpha^{-q}$. To prove the analogue of Theorem~\ref%
{thm-1.10} a different notion of sparsity of potentials is used in \cite%
{RozenblumSolomyak2008, RozenblumSolomyak2009}. In \cite%
{RozenblumSolomyak2010, RozenblumSolomyak2011}, the results are generalized
to fairly arbitrary graphs. Here the interesting observation is made that
the global dimension $D$ defined by the decay $(e^{-t K}(x,x) \leq C \cdot
t^{-D/2}$ of the diagonal elements of the semigroup generated by the kinetic
energy is the quantity that replaces the spatial dimension $d$ of the
hypercubic lattice $\mathbb{Z}^d$. \\[1.5ex]
We give an overview on where to find the proofs of the theorems above:%
\newline

\begin{center}
\begin{tabular}{|p{0.18\textwidth}|c|p{0.68\textwidth}|}
\hline
Theorem~\ref{thm-1.1}: & $\to$ & Theorem~\ref{thm-3.3} of Section~\ref%
{subsec-3.1}. \\ \hline
Theorem~\ref{thm-1.2}: & $\to$ & Corollary~\ref{cor-5.5} of Section \ref%
{sec-5}. \\ \hline
Theorem~\ref{thm-1.3}: & $\to$ & Theorem~\ref{thm-3.5} of Section~\ref%
{subsec-3.2}. \\ \hline
Theorem~\ref{thm-1.4}: & $\to$ & Lemma~\ref{lem-3.6} of Section~\ref%
{subsec-3.2}. \\ \hline
Theorem~\ref{thm-1.5}: & $\to$ & Corollary~\ref{cor-4.5,1} of Section~\ref%
{subsec-4.2}. \\ \hline
Lemma~\ref{lem-1.6}: & $\to$ & Lemma~\ref{lem-4.2} of Section~\ref%
{subsec-4.2}. \\ \hline
Theorems~\ref{thm-1.7} and \ref{thm-1.8}: & $\to$ & Section~\ref{subsec-4.1}.
\\ \hline
Theorems~\ref{thm-1.9}, \ref{thm-liminf-limsup}, and \ref{thm-1.10}: & $\to$
& Corollary~\ref{cor-4.5}, Theorem~\ref{thm-non-sc} and Theorem~\ref{thm-4.6}
of Section~\ref{subsec-4.2}, respectively. \\ \hline
\end{tabular}
\end{center}

\vspace{1.5ex}

As is usual, in order to simplify discussions, some technical results are
proven in the appendix.

\section{Birman-Schwinger Principle and the CLR-Bound}

\label{sec-2}

In the sequel, we use the Birman-Schwinger principle in the following form:

\begin{lemma}[Birman-Schwinger principle]
\label{lem-2.2} Let $d\geq 1$, $\mathfrak{e}$ be an admissible dispersion
relation and $V\in \ell _{0}^{\infty }(\Gamma ,\mathbb{R}_{0}^{+})$. For any 
$\rho >0$, define the compact, self-adjoint, non-negative \textit{%
Birman-Schwinger operator} by 
\begin{equation}
B(\rho )\text{ }=\text{ }B(\rho ,\mathfrak{e},V)\text{ }\doteq \text{ }%
V^{1/2}\,[\rho +h(\mathfrak{e})]^{-1}\,V^{1/2}.  \label{eq-2.3}
\end{equation}%
Then the following assertions~(i)--(iv) hold true.

\begin{enumerate}
\item[(i)] If $\varphi \in \ell^{2}(\Gamma )$ solves $H(\mathfrak{e}%
,V)\varphi =-\rho \varphi $ then $\psi \doteq V^{1/2}\varphi \in \ell
^{2}(\Gamma )$ solves $\psi =B(\rho )\psi $.

\item[(ii)] If $\psi \in \ell^{2}(\Gamma )$ solves $\psi =B(\rho )\psi $
then $\varphi =[\rho +h(\mathfrak{e})]^{-1}V^{1/2}\psi \in \ell^{2}(\Gamma ) 
$ solves $H(\mathfrak{e},V)\varphi =-\rho \varphi $.

\item[(iii)] $-\rho $ is an eigenvalue of $H(\mathfrak{e},V)$ of
multiplicity $M$ if and only if $1$ is an eigenvalue of $B(\rho )$ of
multiplicity $M$.

\item[(iv)] Counting multiplicities, the number of eigenvalues of $H(%
\mathfrak{e},V)$ less or equal than $-\rho $ equals the number of
eigenvalues of $B(\rho )$ greater or equal than $1$.
\end{enumerate}
\end{lemma}

This result is well-know and its proof is given in Appendix \ref{subsec-A.2}
for completeness. The estimate on the number of negative eigenvalues of $H(%
\mathfrak{e},V)$, stated below, is the celebrated CLR bound, which is
generaly derived from (some convenient form of) the Birman-Schwinger
principle.

\begin{theorem}[CLR bound]
\label{thm-2.6} Let $d\geq 3$ and $\mathfrak{e}$ be any admissible
dispersion. Then, for some constant $C_{\ref{thm-2.6}}(d,\mathfrak{e}%
)<\infty $, 
\begin{equation}
N[\mathfrak{e},V]\ \leq \ C_{\ref{thm-2.6}}(d,\mathfrak{e})\,|V|_{d/2}^{d/2}.
\label{eq-2.31}
\end{equation}
\end{theorem}

This kind of estimate was proven the first time by Rozenblum \cite%
{Rozenblum1972}, Lieb \cite{Lieb1976b}, and Cwikel \cite{Cwikel1977} by
three different methods, in the continuous case. See also \cite[%
Theorem~XIII.12]{ReedSimonIV1978} or \cite[Theorem~9.3]{Simon1979a}. It was
then shown by Rozenblum and Solomyak \cite%
{RozenblumSolomyak1997,RozenblumSolomyak2008} that the CLR bound is not only
true for Schr{\"{o}}dinger Operators of the form (\ref{eq-1.1}), but also
for a very large class of operators including, in particular, discrete Schr{%
\"{o}}dinger operators. Note that, when applied to the discrete Schr\"{o}%
dinger operators of the form (\ref{eq-1.8}), most of the known methods to
derive CLR bounds would need the hopping matrix $h(\mathfrak{e})$ to be
positivity preserving. We use instead a beautiful recent observation made by
Frank \cite[Theorem 3.2]{Frank2014} on the discrete spectrum of a class of
selfadjoint operators, which implies the CLR bound for $N[\mathfrak{e},V]$
when $d\geq 3$, merely assuming that $\mathfrak{e}$ is a Morse function
(i.e., it is \textquotedblleft admissible\textquotedblright\ in the sense
defined above). For completeness, in Appendix, Section~\ref{subsec-A.2}, we
reproduce Frank's estimate and derive from it the CLR bound of Theorem \ref%
{thm-2.6}.

\section{Non-Asymptotic Semi-Classical Bounds}

\label{sec-3}

\subsection{Derivation of Non-Asymptotic Bounds}

\label{subsec-3.1}

Now we are in a position to use Theorem~\ref{thm-2.6} to yield a
semi-classical bound, i.e., a bound on $N[\mathfrak{e},V]$ by multiples of $%
N_{sc}[\mathfrak{e},V]$. The following lemma is a standard estimate on the
size of the discrete spectrum of a sum of self-adjoint operators. Its proof
is given in Appendix~\ref{subsec-A.3} for completeness.

\begin{lemma}
\label{lem-3.1} Let $A=A^{\ast },B=B^{\ast }\in \mathcal{B}[\mathcal{H}]$ be
two bounded self-adjoint operators on a separable Hilbert space $\mathcal{H}$%
. Then 
\begin{equation}
N[A+B]\ \leq \ N[A]+N[B]\;,  \label{eq-3.1}
\end{equation}%
where $N[Q]\doteq \mathop{\mathrm{Tr}}\big\{\mathbf{1}[Q<0]\big\}$ denotes
the number of negative eigenvalues, counting multiplicities, of a bounded
self-adjoint operator $Q\in \mathcal{B}[\mathcal{H}]$. We set $N[Q]\doteq
\infty $ if $\sigma _{\mathrm{ess}}(Q)\cap \mathbb{R}^{-}\neq \varnothing $.
\end{lemma}

A simple application of Lemma~\ref{lem-3.1}, with $A\doteq H(\mathfrak{e}%
,V_{1})$, $B=V_{2}$, and $A+B=H(\mathfrak{e},V_{1}+V_{2})$, is the following
corollary.

\begin{corollary}
\label{cor-3.2} Let $d\geq 1$, $\mathfrak{e}$ be an admissible dispersion
relation, and $V_{1},V_{2}\in \ell _{0}^{\infty }(\Gamma ,\mathbb{R}%
_{0}^{+}) $ be two potentials. Then 
\begin{equation}
N[\mathfrak{e},V_{1}+V_{2}]\ \leq \ N[\mathfrak{e},V_{1}]+\#%
\mathop{\mathrm{supp}}\{V_{2}\}.  \label{eq-3.6}
\end{equation}
\end{corollary}

In order to compare the contributions $N[\mathfrak{e},V_{1}]$ and $\#%
\mathop{\mathrm{supp}}\{V_{2}\}$ on the right-hand side of \eqref{eq-3.6} to 
$N_{sc}[\mathfrak{e},V]$, we use the following definition.

\begin{definition}
\label{def-3.3} Let $d\geq 1$. Given a dispersion relation $\mathfrak{e}$
and a potential $V\in \ell _{0}^{\infty }(\Gamma ,\mathbb{R}_{0}^{+})$, we
define: 
\begin{eqnarray}
N_{sc}^{>}[\mathfrak{e},V] &\doteq &\#\big\{x\in \Gamma \,\big|\ V(x)\geq {%
\mathfrak{e}_{\mathrm{max}}}\big\}, \\
N_{sc}^{<}[\mathfrak{e},V] &\doteq &\sum_{x\in \Gamma }\mathbf{1}[V(x)<{%
\mathfrak{e}_{\mathrm{max}}}]\;V^{d/2}(x).
\end{eqnarray}
\end{definition}

Observe that, because dispersion relations are Morse functions, there are
constants $0<c_{1}(\mathfrak{e})\leq c_{2}(\mathfrak{e})<\infty $ such that
for any potential $V\geq 0$,%
\begin{equation}
c_{1}(\mathfrak{e})\Big(N_{sc}^{>}[\mathfrak{e},V]+N_{sc}^{<}[\mathfrak{e},V]%
\Big)\ \leq \ N_{sc}[\mathfrak{e},V]\ \leq \ c_{2}(\mathfrak{e})\Big(%
N_{sc}^{>}[\mathfrak{e},V]+N_{sc}^{<}[\mathfrak{e},V]\Big).  \label{eq-3.9}
\end{equation}

Corollary~\ref{cor-3.2}, \eqref{eq-3.9}, and the CLR bound immediately yield
Theorem~\ref{thm-1.1}:

\begin{theorem}[Thm.~\protect\ref{thm-1.1}]
\label{thm-3.3} Let $d\geq 3$ and $\mathfrak{e}$ be an admissible
dispersion. Then there exists a constant $C_{\ref{thm-3.3}}(d,\mathfrak{e}%
)\in \lbrack 1,\infty )$ such that 
\begin{equation}
N[\mathfrak{e},V]\ \leq \ C_{\ref{thm-3.3}}(d,\mathfrak{e})\,N_{sc}[%
\mathfrak{e},V]\ <\ \infty   \label{eq-3.10}
\end{equation}%
for all $V\in \ell ^{d/2}(\Gamma ,\mathbb{R}_{0}^{+})$.
\end{theorem}

\noindent \textit{Proof}: We apply Corollary~\ref{cor-3.2} to $V=V_{1}+V_{2}$%
, with $V_{1}(x)\doteq V(x)\mathbf{1}[V(x)<${$\mathfrak{e}$}${_{\mathrm{max}}%
}]$ and $V_{2}(x)\doteq V(x)\mathbf{1}[V(x)\geq ${$\mathfrak{e}$}${_{\mathrm{%
max}}}]$, and then Theorem~\ref{thm-2.6} to $N[\mathfrak{e},V_{1}]$. This
gives 
\begin{eqnarray}
N[\mathfrak{e},V] &\leq &N[\mathfrak{e},V_{1}]\,+\,\#\mathop{\mathrm{supp}}%
V_{2}  \label{eq-3.11} \\[1ex]
&\leq &C_{\ref{thm-2.6}}\,N_{sc}^{<}[\mathfrak{e},V]\,+\,N_{sc}^{>}[%
\mathfrak{e},V]\ \;\leq \ \;\frac{C_{\ref{thm-2.6}}+1}{c_{1}(\mathfrak{e})}%
\,N_{sc}[\mathfrak{e},V].
\end{eqnarray}%
{\phantom{A}}\hfill $\square $\bigskip

\subsection{Saturation of the Non-Asymptotic Semi-classical Bounds}

\label{subsec-3.2}

Below, we discuss the optimality of the bound in Theorem~\ref{thm-1.1} in
three different situations: For slowly decaying potentials, for strong and
finitely supported potentials, and for weak potentials which are slowly
varying in space.

We first show that if $V$ decays slower than $|x|^{-2}$ then $0$ is an
accumulation point of the discrete spectrum of $H(\mathfrak{e},V)$ and, in
particular, $H(\mathfrak{e},V)$ has infinitely many negative eigenvalues,
i.e., $N[\mathfrak{e},V]=N_{sc}[\mathfrak{e},V]=\infty $. To formulate the
statement, we recall that $h_{x,y}=h(\mathfrak{e})_{x,y}\doteq \langle
\delta _{x}\,|\,h(\mathfrak{e})\,\delta _{y}\rangle $ denotes matrix
elements of $h(\mathfrak{e})$.

\begin{theorem}[{${N[e,V]}=\infty $ for slowly decaying potentials}]
\label{thm-3.5} Let $\mathfrak{e}$ be an admissible dispersion relation with
hopping matrix $h(\mathfrak{e})$ and $V\in \ell _{0}^{\infty }(\Gamma ,%
\mathbb{R}_{0}^{+})$. Assume that there are constants $\mathrm{const}<\infty 
$ and $\mathrm{const}^{\prime \prime }>0$ with $\alpha <\min \{\alpha
^{\prime },2\}$ such that, for all $x\in \Gamma \backslash \{0\}$, 
\begin{equation}
V(x)\geq \mathrm{const}^{\prime} \cdot |x|^{-\alpha }, \quad |h_{0,x}| \leq 
\mathrm{const} \cdot |x|^{-(2d+\alpha^{\prime })}.  \label{eq-3.16}
\end{equation}
Then $H(\mathfrak{e},V)$ has infinitely many eigenvalues below $0$.
\end{theorem}

The proof of this theorem is a bit lengthy and is given in Appendix~\ref%
{subsec-A.3}. For the case $\mathfrak{e}=\mathfrak{e}_{\mathrm{Lapl}}$ and $%
d=1$. See also \cite{DamanikHundertmarkKillipSimon2003}.

Note that -- assuming $\alpha^{\prime }\geq 2$ -- Theorem~\ref{thm-3.5}
together with the bound \eqref{eq-2.31} implies that the case $V(x)\sim
|x|^{-2}$ is critical in dimension $d\geq 3$ in the sense that 
\begin{align}
\exists \, \varepsilon >0: \ \sup_{x\in \Gamma }\bigg\{\frac{V(x)}{%
|x|^{2+\epsilon }}\bigg\} < \infty \ \ & \Rightarrow \ \ N[\mathfrak{e},V],\
N_{sc}[\mathfrak{e},V]<\infty ,  \label{eq-3.18a} \\[1ex]
\exists \, \varepsilon >0: \ \inf_{x\in \Gamma }\bigg\{\frac{V(x)}{%
|x|^{2-\epsilon }}\bigg\} > 0 \ \ & \Rightarrow \ \ N[\mathfrak{e},V] =
N_{sc}[\mathfrak{e},V] = \infty .  \label{eq-3.18b}
\end{align}
Observe also that Theorem~\ref{thm-1.3} follows from Theorem~\ref{thm-3.5}.

\begin{lemma}[{Lower bound on {$N[e,V]$} without \protect\ref{H-1} and for $%
d\geq 1$}]
\label{lem-3.6} Let $d\geq 1$ and $\mathfrak{e}$ be an admissible dispersion
relation. Furthermore let $V\in \ell _{0}^{\infty }(\Gamma ,\mathbb{R}%
_{0}^{+})$ be a potential decaying at $\infty $. Then, for all $c>\mathfrak{e%
}_{\mathrm{\max }}$, 
\begin{equation}
N[\mathfrak{e},V]\ \geq \ \mathcal{L}_{V}[c]\ =\ \big\{x\in \Gamma \;\big|\
V(x)\geq c\big\}.  \label{eq-3.19}
\end{equation}
\end{lemma}

\noindent \textit{Proof:} For all $\rho >0$, 
\begin{equation}
B(\rho )\ \doteq \ V^{1/2}\frac{1}{\rho +h(\mathfrak{e})}V^{1/2}\geq \frac{1%
}{\mathfrak{e}_{\mathrm{\max }}}V.  \label{eq-3.20}
\end{equation}%
By the min-max principle and Lemma~\ref{lem-2.2} (Birman-Schwinger
principle), we hence obtain that 
\begin{equation}
N[\mathfrak{e},V] \ \geq \ \mathcal{L}_{V}[c] \; ,  \label{eq-3.20,1}
\end{equation}
for all $c > \mathfrak{e}_{\mathrm{\max}}$. \hfill $\square$

\bigskip

The following (stronger) result holds for sparse potentials:

\begin{lemma}[{Lower bound on {$N[e,V]$} \ for sparse potentials}]
\label{lem-3.6.b} Let $d\geq 3$ and $\mathfrak{e}$ be an admissible
dispersion relation. Let $0<\eta (\mathfrak{e})<$ $\mathfrak{e}_{\mathrm{%
\max }}$ be defined by 
\begin{equation*}
\frac{1}{\eta (\mathfrak{e})}=\int [\mathfrak{e}(p)]^{-1}\,\mathrm{d}\mu
^{\ast }(p).
\end{equation*}%
Furthermore, let $V\in \ell _{0}^{\infty }(\Gamma ,\mathbb{R}_{0}^{+})$ be a
potential which is sparse in the sense that 
\begin{equation*}
\eta(\mathfrak{e}) \cdot \sup_{\rho >0} \bigg\{ \sup_{x\in \mathrm{supp\;}V} %
\bigg( \sum_{y\in \mathrm{supp\;}V\setminus \{x\}} \big| \langle \delta _{x}
| \; [\rho +h(\mathfrak{e})]^{-1} \delta _{y}\rangle \big| \bigg) \bigg\} \
< \ \frac{\varepsilon }{1+\varepsilon } \ < \ 1 \; ,
\end{equation*}
for some $0<\varepsilon <\infty $. Then 
\begin{equation}
N[\mathfrak{e},V]\ \geq \ \mathcal{L}_{V}[(1+\varepsilon )\eta (\mathfrak{e}%
)]\ =\ \big\{x\in \Gamma \;\big|\ V(x)\geq (1+\varepsilon )\eta (\mathfrak{e}%
)\big\}.  \label{eq-3.19.b}
\end{equation}
\end{lemma}

\noindent \textit{Proof:} Observe that $N[\mathfrak{e},V]\geq N[\mathfrak{e}%
,V^{\prime }]$ with $V^{\prime }(x)\doteq \max \{V(x),(1+\varepsilon )\eta (%
\mathfrak{e})\}$. Let $\rho >0$ and $x\in \Gamma $. Similarly to %
\eqref{eq-2.26}, we have 
\begin{equation}
\langle \delta _{x}|\,B(\rho ,\mathfrak{e},V^{\prime })\delta _{x}\rangle \
=\ V^{\prime }(x)\bigg(\int_{\Gamma ^{\ast }}\frac{\mathrm{d}\mu ^{\ast }(p)%
}{\rho +\mathfrak{e}(p)}\bigg).
\end{equation}%
Observe that, by the assumption on $V$ and the Schur bound, for all $\psi
\in \ell ^{2}(\Gamma )$ ,%
\begin{equation*}
\sup_{\rho >0}\langle \psi |\,B(\rho ,\mathfrak{e},V^{\prime })\psi \rangle
\ >\ \bigg(\sum_{x\in \mathcal{L}}|\psi _{x}|^{2}(1+\varepsilon )\bigg)%
-\varepsilon .
\end{equation*}%
where the summation runs over $x\in \mathcal{L}\doteq \mathcal{L}%
_{V}[(1+\varepsilon )\eta (\mathfrak{e})]=\big\{x\in \Gamma \big|\,V(x)\geq
(1+\varepsilon )\eta (\mathfrak{e})\big\}$. By Lemma~\ref{lem-2.2}
(Birman-Schwinger principle) and the min-max principle, we hence obtain that 
\begin{equation}
N[\mathfrak{e},V^{\prime }]\ \geq \ \mathcal{L}_{V}[(1+\varepsilon )\eta (%
\mathfrak{e})].\hspace{5mm}
\end{equation}%
{\phantom{A}}\hfill $\square $\bigskip

Note that Lemma~\ref{lem-3.6}, together with Corollary~\ref{cor-3.2} and $N[%
\mathfrak{e},0]=0$, implies that, for finitely supported potentials $V$, we
have 
\begin{equation}
\lim_{\lambda \rightarrow \infty }N[\mathfrak{e},\lambda V]=\lim_{\lambda
\rightarrow \infty }N_{sc}[\mathfrak{e},\lambda V]=\mathop{\mathrm{supp}}V,
\label{eq-3.21}
\end{equation}%
and thus the semi-classical upper bound on $N[\mathfrak{e},\lambda V]$
saturates when $\lambda \rightarrow \infty $.

Observe further that, on one hand, Theorem~\ref{w.s.var.pot} below implies
that the lower bound on $N[\mathfrak{e},V]$ given in Lemma~\ref{lem-3.6}
strongly underestimates the size of the discrete spectrum of $H(\mathfrak{e}%
,V)$ in the case where $V$ is slowly varying in space. $N_{sc}[\mathfrak{e}%
,V]$ describes -- in this precise case -- the behavior of $N[\mathfrak{e},V]$
more correctly. On the other hand, it seems that there is no other simple
candidate for a lower bound on $N[\mathfrak{e},V]$ holding in general and
based on quantities like \ $N_{sc}[\mathfrak{e},V]$ or $|V|_{p}^{p}$. See
Corollary~\ref{cor-4.5} and remark thereafter.

For any continuous function $f:{\mathbb{R}}^{d}\rightarrow {\mathbb{R}}%
_{0}^{+}$ define for all $M\in {\mathbb{N}}_{0}$ the step functions $%
f_{-}^{(M)}:{\mathbb{R}}^{d}\rightarrow {\mathbb{R}}_{0}^{+}$ by: 
\begin{equation}
f_{-}^{(M)}(x)\doteq \sum\limits_{X\in {\mathbb{Z}}^{d}}\mathbf{1}[x\in
2^{-M}X+[0,2^{-M})^{d}]\min \{\,f(x^{\prime })\,|\,x^{\prime }\in
2^{-M}X+[0,2^{-M})^{d}\}.
\end{equation}

\begin{lemma}
\label{l.b.cont} Let $v\in C_{0}({\mathbb{R}}^{d},{\mathbb{R}}_{0}^{d})$ be
compactly supported. For all $L>0$ define the potential $V_{L}:\Gamma
\rightarrow {\mathbb{R}}_{0}^{+}$ by: 
\begin{equation}
V_{L}(x)\doteq L^{-2}v(L^{-1}x).
\end{equation}%
Let $\mathfrak{e}$ be any admissible dispersion relation from $C^{3}(\Gamma
^{\ast },{\mathbb{R}})$. Assume, moreover, that for some $D<\infty $ and
some $\alpha >2$, for all $x\in \Gamma $, 
\begin{equation}
|h(\mathfrak{e})_{0,x}|\leq D\,\langle x\rangle^{2d+\alpha }.
\end{equation}%
Then there are constants $\mathrm{const}^{\prime }>0$, $\mathrm{const}%
<\infty $, depending only $\mathfrak{e}$ such that for all $M\in {\mathbb{N}}%
_{0}$, 
\begin{equation}
\liminf_{L\rightarrow \infty }N[\mathfrak{e},V_{L}]\geq \mathrm{const}%
^{\prime }\,\int_{{\mathbb{R}}^{d}}\,v_{-}^{(M)}(x)^{d/2}\,\mathbf{1}%
[v_{-}^{(M)}(x)>\mathrm{const}^{2M}]\mathrm{d}^{d}x.
\label{lim.inf.continuum}
\end{equation}
\end{lemma}

We prove this by standard arguments using coherent states. See Appendix \ref%
{subsec-A.3}. The following result is an immediate consequence of the lemma
above.

\begin{theorem}
\label{w.s.var.pot} Let $\mathfrak{e}$ be any admissible dispersion relation
from $C^{3}(\Gamma^{\ast },{\mathbb{R}})$ and $v\in C_{0}({\mathbb{R}}^{d},{%
\mathbb{R}}_{0}^{d})$ be compactly supported. Let the potentials $%
V_{L}=V_{L}(v)$ be defined as above. Then, for some constant $\mathrm{const}%
>0$ depending only on $\mathfrak{e}$, 
\begin{equation}
\liminf_{\lambda \rightarrow \infty }\liminf_{L\rightarrow \infty }N[%
\mathfrak{e},\lambda V_{L}]\geq \mathrm{const}\,\lambda^{d/2}\int_{{\mathbb{R%
}}^{d}}\,v(x)^{d/2}\mathrm{d}^{d}x.
\end{equation}
\end{theorem}

Observe, moreover, that from Theorem~\ref{w.s.var.pot}: $N[\mathfrak{e}%
,\lambda V_{L}]\geq \mathrm{const\;}N_{sc}[\mathfrak{e},\lambda V_{L}]$ for
some $\mathrm{const}>0$ and sufficiently large $\lambda >0$ and $L>0$. Thus,
as expected, like in the continuous case: $N[\mathfrak{e},\lambda V_{L}]\sim
N_{sc}[\mathfrak{e},\lambda V_{L}]$ at large $\lambda >0$ and $L>0$.

\section{Asymptotics of ${N[}${$\mathfrak{e}$}${,\protect\lambda V]}$ for
large $\protect\lambda $}

\label{sec-4}

In this section we investigate the question whether the semi-classical
number of bound states $N_{sc}[\mathfrak{e},\lambda V]$ describes $N[%
\mathfrak{e},\lambda V]$ correctly in the limit $\lambda \rightarrow \infty $
or not. This leads us to the proof of Theorems~\ref{thm-1.7} and \ref%
{thm-1.8}.

Equally interesting, however, is the observation made in this section that
an asymptotic comparison of $N[\mathfrak{e},\lambda V]$ to $N_{sc}[\mathfrak{%
e},\lambda V]$ does not always make much sense. Namely, in Theorem~\ref%
{thm-4.6} below, we prove that $\lambda \mapsto N[\mathfrak{e},\lambda V]$
may approximate any given continuous and monotonically increasing function $%
F(\lambda )$ of $\lambda $. More precisely, given $F$, we can always find a
potential $V_{F}$ such that $N[\mathfrak{e},\lambda V_{F}]=F(\lambda )$ up
to a small error.

\subsection{ Potentials with Semi-classical Asymptotic \newline
Behavior of $N{[}${$\mathfrak{e}$}${,\protect\lambda V]}$ at large $\protect%
\lambda $}

\label{subsec-4.1}

This subsection is devoted to the proof of Theorems~\ref{thm-1.7} and \ref%
{thm-1.8}. To this end, we recall that 
\begin{eqnarray}
g_{+}(V) &\doteq &\sup_{r>0}\limsup_{\ell \rightarrow \infty }\frac{2}{d\,r}%
\Big(\ln \mathcal{L}_{V}\big[\mathrm{e}^{-\ell -r}\big]-\ln \mathcal{L}_{V}%
\big[\mathrm{e}^{-\ell }\big]\Big),  \label{eq-4.2} \\[1ex]
g_{-}(V) &\doteq &\inf_{r>0}\liminf_{\ell \rightarrow \infty }\frac{2}{d\,r}%
\Big(\ln \mathcal{L}_{V}\big[\mathrm{e}^{-\ell -r}\big]-\ln \mathcal{L}_{V}%
\big[\mathrm{e}^{-\ell }\big]\Big).
\end{eqnarray}%
The following lemma illustrates that, for potentials with $g_{+}(V)<1$, the
main contribution to $N_{sc}[\mathfrak{e},\lambda V]$ is given by $%
\#\{\lambda V\geq \,${$\mathfrak{e}$}${_{\mathrm{max}}}\}$, and that this
actually defines a borderline in the sense that if $g_{-}(V)\geq 1$ then
this assertion is reversed.

\begin{lemma}
\label{lem-4.1} Assume $d\geq 1$ and $V\in \ell _{0}^{\infty }(\Gamma ,%
\mathbb{R}_{0}^{+})$.

\begin{itemize}
\item[(i)] Then there is a constant $C_{\ref{lem-4.1}}(d,\mathfrak{e})>0$
such that 
\begin{equation}
\liminf_{\lambda \rightarrow \infty }\bigg\{\frac{N_{sc}^{>}[\mathfrak{e}%
,\lambda V]}{N_{sc}[\mathfrak{e},\lambda V]}\bigg\}\ \geq \ \big(1-g_{+}(V)%
\big)\,C_{\ref{lem-4.1}}(d,\mathfrak{e}).  \label{eq-4.3}
\end{equation}

\item[(ii)] Conversely, if $g_{-}(V)\geq 1$ then 
\begin{equation}
\lim_{\lambda \rightarrow \infty }\bigg\{\frac{N_{sc}^{>}[\mathfrak{e}%
,\lambda V]}{N_{sc}[\mathfrak{e},\lambda V]}\bigg\}\ =\ 0,  \label{eq-4.4}
\end{equation}%
where $N_{sc}^{>}[\mathfrak{e},V]=\mathcal{L}_{V}[${$\mathfrak{e}$}${_{%
\mathrm{max}}}]=\#\{V\geq \,${$\mathfrak{e}$}${_{\mathrm{max}}}\}$ is
defined in Definition~\ref{def-3.3}.
\end{itemize}
\end{lemma}

\noindent \textit{Proof:} We first fix $x\in \Gamma $, set $\rho _{x}\doteq
\min \big\{1,\;\lambda V(x)/${$\mathfrak{e}$}${_{\mathrm{max}}}\big\}$, and
observe that 
\begin{equation}
c_{1}\,\rho _{x}^{d/2}\ \leq \ \int_{\Gamma^{\ast }}\mathbf{1}[\mathfrak{e}%
(p)<\lambda V(x)]\mathrm{d}\mu^{\ast }(p)\ \leq \ C_{1}\,\rho _{x}^{d/2},
\label{eq-4.5}
\end{equation}%
for some $0<c_{1}\equiv c_{1}(d,\mathfrak{e})<C_{1}\equiv C_{1}(d,\mathfrak{e%
})<\infty $, since $\mathfrak{e}(p)$ is a Morse function. Furthermore, we
have that 
\begin{eqnarray}
N_{sc}[\mathfrak{e},\lambda V] &=&\sum_{x\in \Gamma }\int_{\Gamma^{\ast }}%
\mathbf{1}[\mathfrak{e}(p)<\lambda V(x)]\,\mathrm{d}\mu^{\ast }(p)
\label{eq-4.6} \\[1ex]
&=&\sum_{x\in \Gamma }\int_{\Gamma^{\ast }}\mathbf{1}[\mathfrak{e}%
(p)<\lambda V(x)\leq {\mathfrak{e}_{\mathrm{max}}}]\,\mathrm{d}\mu^{\ast
}(p)+\mathcal{L}_{V}[\lambda^{-1}{\mathfrak{e}_{\mathrm{max}}}].  \notag
\end{eqnarray}%
Using that 
\begin{equation}
\rho _{x}^{d/2}\ =\ \frac{d}{2}\int_{0}^{\infty }\mathbf{1}\big[\mathrm{e}%
^{-r}<\rho _{x}\big]\,\mathrm{e}^{-dr/2}\,\mathrm{d}r  \label{eq-4.6,1}
\end{equation}%
and $\ell _{\lambda }\doteq \log (\lambda )-\log (${$\mathfrak{e}$}${_{%
\mathrm{max}}})$, we hence obtain 
\begin{eqnarray}
\lefteqn{N_{sc}[\mathfrak{e},\lambda V]-\cL_{V}\big\lbrack \mathrm{e}^{-\ell
_{\lambda }}\big]\ }  \notag  \label{eq-4.7} \\[1ex]
\; &=&\sum_{x\in \Gamma }\int_{\Gamma^{\ast }}\mathbf{1}[\mathfrak{e}%
(p)<\lambda V(x)<{\mathrm{e}_{\mathrm{max}}}]\mathrm{d}\mu^{\ast }(p) \\
&\leq &\frac{dC_{1}}{2}\sum_{x\in \Gamma }\int_{0}^{\infty }\Big\{\mathbf{1}%
\big[\mathrm{e}^{-r}\leq \lambda {\mathfrak{e}_{\mathrm{max}}^{-1}}V(x)\big]-%
\mathbf{1}\big[1\leq \lambda {\mathfrak{e}_{\mathrm{max}}^{-1}}V(x)\big]%
\Big\}\,\mathrm{e}^{-dr/2}\,\mathrm{d}r,  \notag \\[1ex]
&=&\frac{dC_{1}}{2}\int_{0}^{\infty }\Big\{\mathcal{L}_{V}\big[\mathrm{e}%
^{-\ell _{\lambda }-r}\big]-\mathcal{L}_{V}\big[\mathrm{e}^{-\ell _{\lambda
}}\big]\Big\}\,\mathrm{e}^{-dr/2}\,\mathrm{d}r,  \notag \\[1ex]
&=&\frac{d\,C_{1}\,\mathcal{L}_{V}[\mathrm{e}^{-\ell _{\lambda }}]}{2}%
\int_{0}^{\infty }\bigg\{\frac{\mathcal{L}_{V}[\mathrm{e}^{-\ell _{\lambda
}-r}]}{\mathcal{L}_{V}[\mathrm{e}^{-\ell _{\lambda }}]}\bigg\}\,\mathrm{e}%
^{-dr/2}\,\mathrm{d}r-C_{1}\,\mathcal{L}_{V}\big[\mathrm{e}^{-\ell _{\lambda
}}\big].
\end{eqnarray}%
and similarly 
\begin{eqnarray}
&&N_{sc}[\mathfrak{e},\lambda V]-\mathcal{L}_{V}\big[\mathrm{e}^{-\ell
_{\lambda }}\big]\  \\
&\geq &\ \frac{d\,c_{1}\,\mathcal{L}_{V}[\mathrm{e}^{-\ell _{\lambda }}]}{2}%
\int_{0}^{\infty }\bigg\{\frac{\mathcal{L}_{V}[\mathrm{e}^{-\ell _{\lambda
}-r}]}{\mathcal{L}_{V}[\mathrm{e}^{-\ell _{\lambda }}]}\bigg\}\,\mathrm{e}%
^{-dr/2}\,\mathrm{d}r-c_{1}\,\mathcal{L}_{V}\big[\mathrm{e}^{-\ell _{\lambda
}}\big].  \notag
\end{eqnarray}%
Defining 
\begin{equation}
g_{\ell }(r)\ \doteq \ \frac{2}{d\,r}\Big(\ln \mathcal{L}_{V}\big[\mathrm{e}%
^{-\ell -r}\big]-\ln \mathcal{L}_{V}\big[\mathrm{e}^{-\ell }\big]\Big),
\label{eq-4.9}
\end{equation}%
we hence have 
\begin{eqnarray}
\frac{d\,C_{1}}{2}\int_{0}^{\infty }\exp \left( -[1-g_{\ell _{\lambda
}}(r)]\,\frac{d}{2}r\right) \,\mathrm{d}r &\geq &\frac{N_{sc}[\mathfrak{e}%
,\lambda V]}{\mathcal{L}_{V}[\mathrm{e}^{-\ell _{\lambda }}]}-1+C_{1}.
\label{eq-4.11} \\[1ex]
\frac{d\,c_{1}}{2}\int_{0}^{\infty }\exp \left( -[1-g_{\ell _{\lambda
}}(r)]\,\frac{d}{2}r\right) \,\mathrm{d}r &\leq &\frac{N_{sc}[\mathfrak{e}%
,\lambda V]}{\mathcal{L}_{V}[\mathrm{e}^{-\ell _{\lambda }}]}-1+c_{1},
\end{eqnarray}%
Now, an application of Fatou's Lemma yields 
\begin{eqnarray}
\limsup_{\lambda \rightarrow \infty }\frac{N_{sc}[\mathfrak{e},\lambda V]}{%
\mathcal{L}_{V}[\mathrm{e}^{-\ell _{\lambda }}]} &\leq &1-C_{1}+\frac{%
d\,C_{1}}{2}\int_{0}^{\infty }\exp \left( -[1-g_{+}(V)]\right) \,r\mathrm{d}r
\notag  \label{eq-4.12} \\[1ex]
&=&1-C_{1}+\frac{dC_{1}}{[1-g_{+}(V)]},
\end{eqnarray}%
which implies (i). Assertion~(ii) is similar, for if $g_{-}(V)\geq 1$ then
another application of Fatou's Lemma gives 
\begin{equation}
\liminf_{\lambda \rightarrow \infty }\frac{N_{sc}[\mathfrak{e},\lambda V]}{%
\mathcal{L}_{V}[\mathrm{e}^{-\ell _{\lambda }}]}\ \geq \ 1-c_{1}+\frac{%
d\,c_{1}}{2}\int_{0}^{\infty }\exp \left( [g_{-}(V)-1]\,\frac{d}{2}r\right)
\,\mathrm{d}r\ =\ \infty .  \label{eq-4.13}
\end{equation}%
{\phantom{A}}\hfill $\square $\bigskip

\noindent \textbf{Proof of Theorems~\ref{thm-1.7} and \ref{thm-1.8}:} By
Theorem~\ref{thm-1.4} and Definition~\ref{def-3.3}, we have 
\begin{equation}
\frac{N[\mathfrak{e},\lambda V]}{N_{sc}[\mathfrak{e},\lambda V]}\ \geq \ 
\frac{\mathcal{L}_{V}[\lambda^{-1}{\mathfrak{e}_{\mathrm{max}}}]}{N_{sc}[%
\mathfrak{e},\lambda V]}\ =\ \frac{N_{sc}^{>}[\mathfrak{e},\lambda V]}{%
N_{sc}[\mathfrak{e},\lambda V]}.  \label{eq-4.14}
\end{equation}%
Now, the left-hand inequality in ~\eqref{eq-1.31} and the first inequality
in \eqref{eq-1.33} follow directly from Lemma~\ref{lem-4.1}~(i). The
right-hand inequality in ~\eqref{eq-1.31} follows from Theorem~\ref{thm-1.1}%
, while the second inequality in~\eqref{eq-1.33} is a consequence of Theorem~%
\ref{thm-1.2}.\newline
{\phantom{A}} \hfill $\square $

\subsection{Failure of Semi-classical Asymptotic \newline
Behavior of $N{[}${$\mathfrak{e}$}${,\protect\lambda V]}$ at large $\protect%
\lambda $}

\label{subsec-4.2}

For the continuum Schr{\"{o}}dinger operator $-\Delta -\lambda V(x)$ on $%
\mathbb{R}^{d}$, the number of negative eigenvalues is asymptotically
homogeneous of degree $d/2$ in $\lambda $, i.e., $N_{sc}^{cont}[\lambda
V]=\lambda^{d/2}\,N_{sc}^{cont}[V]$. For discrete Schr\"{o}dinger operators,
only weaker statements hold true, as is illustrated by the following lemma.
See also \cite[Section~5.2]{RozenblumSolomyak2009}.

\begin{lemma}[Lemma~\protect\ref{lem-1.6}]
\label{lem-4.2} Assume $d\geq 3$ and $V\in \ell ^{d/2}(\Gamma ,\mathbb{R}%
_{0}^{+})$. Then 
\begin{equation}
\lim_{\lambda \rightarrow \infty }\big\{\lambda ^{-d/2}\,N[\mathfrak{e}%
,\lambda V]\big\}\ =\ \lim_{\lambda \rightarrow \infty }\big\{\lambda
^{-d/2}\,N_{sc}[\mathfrak{e},\lambda V]\big\}\ =\ 0.  \label{eq-4.15}
\end{equation}
\end{lemma}

\noindent \textit{Proof:} It suffices to prove the second equality, since $N[%
\mathfrak{e},\lambda V]\leq C_{\ref{thm-2.6}}(d,\mathfrak{e})N_{sc}[%
\mathfrak{e},\lambda V]$, by Theorem~\ref{thm-1.1}. By \eqref{eq-3.9}, we
have that 
\begin{equation}
\lambda^{-d/2}\,N[\mathfrak{e},\lambda V]\ \leq \ c_{2}(\mathfrak{e}%
)\,\lambda^{-d/2}\,\Big(N_{sc}^{>}[\mathfrak{e},\lambda V]+N_{sc}^{<}[%
\mathfrak{e},\lambda V]\Big),  \label{eq-4.16}
\end{equation}%
and 
\begin{eqnarray}
\lefteqn{\lambda^{-d/2}\left( N_{sc}^{>}[\mathfrak{e},\lambda V]+N_{sc}^{<}[%
\mathfrak{e},\lambda V]\right) }  \label{eq-4.17} \\[1ex]
&=&\lambda^{-d/2}\,\sum_{x\in \Gamma }\min \big\{{\mathfrak{e}_{\mathrm{max}}%
},\lambda^{d/2}\,V^{d/2}(x)\big\}\ =\ \sum_{x\in \Gamma }\min \big\{%
\lambda^{-d/2}\,{\mathfrak{e}_{\mathrm{max}}},V^{d/2}(x)\big\}.  \notag
\end{eqnarray}%
Since 
\begin{equation*}
\lim_{\lambda \rightarrow \infty }\min \{\lambda^{-d/2}\, \mathfrak{e}_{%
\mathrm{max}},V^{d/2}(x)\}=0
\end{equation*}
for every $x\in \Gamma $ and $\min \{\lambda^{-d/2}\,${$\mathfrak{e}$}${_{%
\mathrm{max}}},V^{d/2}\}$ is dominated by $V^{d/2}\in \ell^{1}(\Gamma )$,
the assertion follows from the dominated convergence theorem. \hfill $%
\square $

\begin{lemma}
\label{lem-4.3} Let $d\geq 3$ and $\mathfrak{e}$ be an admissible dispersion
relation. Then there is a constant $C_{\ref{lem-4.3}}(d,\mathfrak{e})<\infty 
$ such that, for all $\rho \in (0,1]$ and all $x,y\in \Gamma $, $x\neq y$, 
\begin{equation}
\big|\big\la\delta _{x}\,\big|\;\big(\rho +h(\mathfrak{e})\big)^{-1}\delta
_{y}\big\ra\big|\ \leq \ \frac{C_{\ref{lem-4.3}}(d,\mathfrak{e})}{|x-y|^{1/2}%
}.  \label{eq-4.18}
\end{equation}
\end{lemma}

\noindent \textit{Proof:} Let $\mathrm{Min}(\mathfrak{e})\doteq \{\xi \in
\Gamma^{\ast }\,|\,\mathfrak{e}(\xi )=0\}$ be the set of points in $\Gamma
^{\ast }$ for which $\mathfrak{e}$ is minimal. We construct a partition of
unity localizing on the Voronoi cells 
\begin{equation}
\mathcal{V}(\xi )\ \doteq \ \big\{p\in \Gamma^{\ast }\,\big|\,\gamma (p,\xi
)=\min_{{\tilde{\xi}}\in \mathrm{Min}(\mathfrak{e})}\gamma (p,{\tilde{\xi}})%
\big\},  \label{eq-4.19}
\end{equation}%
where $\xi \in \mathrm{Min}(\mathfrak{e})$ and $\gamma :\Gamma^{\ast }\times
\Gamma^{\ast }\rightarrow \mathbb{R}_{0}^{+}$ is the natural metric on $%
\Gamma^{\ast }=(\mathbb{R}/2\pi \mathbb{Z)}^{d}$. Denote by $r>0$ the
largest radius, such that $B_{\gamma }(\xi ,2r)\subseteq \mathcal{V}(\xi )$,
for all $\xi \in \mathrm{Min}(\mathfrak{e})$, and choose $j\in C_{0}^{\infty
}(\mathbb{R}^{d},\mathbb{R}_{0}^{+})$ such that $\mathop{\mathrm{supp}}%
j\subseteq B(0,1)$ and $\int_{\mathbb{R}^{d}}j(p)\,\mathrm{d}^{d}p=1$. We
then set $j_{r}(p)\doteq r^{-d}j(p/r)$ for $p\in \Gamma^{\ast }$ (which
makes sense because $r>0$ is sufficiently small), and 
\begin{equation}
\chi _{\xi }\ \doteq \ j_{r}\ast \mathbf{1}_{\mathcal{V}(\xi )}.
\label{eq-4.20}
\end{equation}%
We list a few properties of this partition in combination with the
dispersion $\mathfrak{e}$ deriving from the fact that $\mathfrak{e}$ is a
Morse function. 
\begin{gather}
\forall \, {p\in \Gamma^{\ast }}:\;\sum_{\xi \in \mathrm{Min}(\mathfrak{e}%
)}\chi _{\xi }(p)=\ 1,  \label{eq-4.25} \\
\forall \, {p\in \Gamma^{\ast }} \,\forall \, {\xi ,{\tilde{\xi}}\in \mathrm{%
Min}(\mathfrak{e}),\,\xi \neq {\tilde{\xi}}}:\;\chi _{\xi
}(p)>0\Longrightarrow \gamma (p,{\tilde{\xi}})>r,  \notag \\[1ex]
\exists _{c_{1}>0}\ \forall \, {p\in \Gamma^{\ast }}\forall \, {\xi \in 
\mathrm{Min}(\mathfrak{e})}:\;\nabla _{p}\chi _{\xi }(p)>0\Longrightarrow 
\mathfrak{e}(p)\geq c_{1},  \notag \\[1ex]
\exists _{c_{2}>0}\ \forall \, {p\in \Gamma^{\ast }}\forall \, {\xi \in 
\mathrm{Min}(\mathfrak{e})}:\;\chi _{\xi }(p)>0\Longrightarrow \mathfrak{e}%
(p)\geq c_{2}(p-\xi )^{2},  \notag \\[1ex]
\exists _{c_{3}<\infty }\ \forall \, {p\in \Gamma^{\ast }}\forall \, {\xi
\in \mathrm{Min}(\mathfrak{e})}:\;\chi _{\xi }(p)>0\Longrightarrow |\nabla 
\mathfrak{e}(p)|\leq c_{3}|p-\xi |.  \notag
\end{gather}%
By translation invariance, it suffices to prove \eqref{eq-4.18} for $y=0$
and $x\neq 0$. We observe that 
\begin{eqnarray}
\lefteqn{|x|^{2}\;\big|\big\la\delta _{x}\,\big|\;\big(\rho +h(\mathfrak{e}%
)\big)^{-1}\delta _{0}\big\ra\big|\ \;=\ \;\bigg|\int_{\Gamma^{\ast }}\frac{%
x\cdot \nabla _{p}\big(e^{ip\cdot x}\big)\;d\mu^{\ast }(p)}{\rho +\mathfrak{e%
}(p)}\bigg|}  \label{eq-4.26} \\[1ex]
&=&\bigg|\sum_{\xi \in \mathrm{Min}(\mathfrak{e})}\int_{\Gamma^{\ast
}}x\cdot \nabla _{p}\big(e^{i(p-\xi )\cdot x}-1\big)\;\frac{\chi _{\xi
}(p)\;d\mu^{\ast }(p)}{\rho +\mathfrak{e}(p)}\bigg|  \notag \\[1ex]
&=&\bigg|\sum_{\xi \in \mathrm{Min}(\mathfrak{e})}\int_{\Gamma^{\ast }}\big(%
e^{i(p-\xi )\cdot x}-1\big)\bigg\{\frac{x\cdot \nabla _{p}\chi _{\xi }(p)}{%
\rho +\mathfrak{e}(p)}\,-\,\frac{\chi _{\xi }(p)\,x\cdot \nabla _{p}%
\mathfrak{e}(p)}{[\rho +\mathfrak{e}(p)]^{2}}\bigg\}\,d\mu^{\ast }(p)\bigg|.
\notag
\end{eqnarray}%
Now we use \eqref{eq-4.25}, $|\mathrm{e}^{i(p-\xi )\cdot x}-1|\leq 2$, and $|%
\mathrm{e}^{i(p-\xi )\cdot x}-1|\leq 2\,|x|^{1/2}\,|p-\xi |^{1/2}$ to obtain 
\begin{eqnarray}
\lefteqn{|x|^{1/2}\,\big|\big\la\delta _{x}\,\big|\;\big(\rho +h(\mathfrak{e}%
)\big)^{-1}\delta _{0}\big\ra\big|}  \label{eq-4.27} \\[1ex]
&\leq &\sum_{\xi \in \mathrm{Min}(\mathfrak{e})}\int_{\Gamma^{\ast }}\bigg\{%
\frac{2\,|\nabla _{p}\chi _{\xi }(p)|}{c_{1}}\,+\,\frac{\chi _{\xi
}(p)\,c_{3}}{c_{2}^{2}(p-\xi )^{5/2}}\bigg\}\,\mathrm{d}\mu^{\ast }(p)\
\;\leq \ \;C_{4},  \notag
\end{eqnarray}%
for some constant $C_{4}<\infty $, since $|p-\xi |^{-5/2}$ is locally
integrable for $d\geq 3$. We remark that we may have improved this estimate
to $\mathcal{O}(|x|^{\beta -1})$, for any $\beta >0$, by using $|\mathrm{e}%
^{i(p-\xi )\cdot x}-1|\leq 2\,|x|^{\beta }\,|p-\xi |^{\beta }$. \hfill $%
\square $

\begin{lemma}
\label{lem-4.4} Let $d\geq 3$ and $\mathfrak{e}$ be an admissible
dispersion. Let ${\underline{r}}\doteq (r_{k})_{k=0}^{\infty }$ be an
increasing sequence of positive integers with $9r_{k}\leq r_{k+1}$ for all $%
k\geq 0$, and define $\omega ({\underline{r}})\doteq
\{x_{0},x_{1},x_{2},\ldots \}\subseteq \Gamma $ by 
\begin{equation}
x_{k}\ \doteq \ (r_{k},0,\ldots ,0).  \label{eq-4.28}
\end{equation}%
If $V\in \ell^{\infty }(\Gamma )$ with $\mathop{\mathrm{supp}}V\subseteq
\omega ({\underline{r}})$ and 
\begin{equation}
|V|_{\infty }\ <\ \eta (\mathfrak{e})-\frac{1}{4}\,C_{\ref{lem-4.3}}(d,%
\mathfrak{e})\,\eta (\mathfrak{e})^{2}\,r_{0}^{-1/2},  \label{eq-4.29}
\end{equation}%
then $N[\mathfrak{e},V]=0$.
\end{lemma}

\noindent \textit{Proof:} For any normalized $\psi =(\psi _{x})_{x\in \Gamma
}\in \ell^{2}(\Gamma )$ and all $\rho >0$, we have that 
\begin{eqnarray}
\lefteqn{\big\la\psi \,\big|\ V^{1/2}\,(\rho +h(\mathfrak{e}%
))^{-1}\,V^{1/2}\psi \big\ra\ }  \label{eq-4.30} \\[1ex]
\; &\leq &\ \;\frac{1}{\eta (\mathfrak{e})}\sum_{x\in \omega ({\underline{r}}%
)}V(x)\,|\psi _{x}|^{2} \\
&&\;+\;\sum\limits_{x,y\in \omega ({\underline{r}}),\;x\neq y}\overline{\psi
_{x}}\psi _{y}\,[V(x)V(y)]^{1/2}\,\big\la\delta _{x}\,\big|\;\big(\rho +h(%
\mathfrak{e})\big)^{-1}\delta _{y}\big\ra  \notag \\[1ex]
&\leq &|V|_{\infty }\,\bigg(\frac{1}{\eta (\mathfrak{e})}\;+\;\sup_{x\in
\omega ({\underline{r}})}\bigg\{\sum_{y\in \omega ({\underline{r}})\setminus
\{x\}}\big|\big\la\delta _{x}\,\big|\;\big(\rho +h(\mathfrak{e})\big)%
^{-1}\delta _{y}\big\ra\big|\bigg\}\bigg),  \notag
\end{eqnarray}%
by the Schur bound. From Lemma~\ref{lem-4.3} it follows that 
\begin{equation}
\sup_{x\in \omega ({\underline{r}})}\bigg\{\sum_{y\in \omega ({\underline{r}}%
)\setminus \{x\}}\big|\big\la\delta _{x}\,\big|\;\big(\rho +h(\mathfrak{e})%
\big)^{-1}\delta _{y}\big\ra\big|\bigg\}\bigg)\ \leq \ C_{\ref{lem-4.3}}(d,%
\mathfrak{e})\,\sup_{k\geq 0}\big\{X_{k}+Y_{k}\big\},  \label{eq-4.31}
\end{equation}%
where 
\begin{equation}
X_{k}\ \doteq \ \sum_{\ell =0}^{k-1}|r_{k}-r_{\ell }|^{-1/2}\quad \text{and}%
\quad Y_{k}\ \doteq \ \sum_{\ell =k+1}^{\infty }|r_{k}-r_{\ell }|^{-1/2}.
\label{eq-4.32}
\end{equation}%
For $\ell <k$, we have that $|r_{k}-r_{\ell }|\geq 8r_{k}\geq 8\cdot
9^{k}\,r_{0}$ and hence 
\begin{equation}
X_{k}\ \leq \ \frac{k\,3^{-k}}{\sqrt{8\,r_{0}}}.  \label{eq-4.33}
\end{equation}%
Similarly, we have that $|r_{k}-r_{\ell }|\geq 8r_{\ell }\geq 8\cdot 9^{\ell
}\,r_{0}$ for $\ell >k$, and thus 
\begin{equation}
Y_{k}\ \leq \ \frac{3^{-k}}{3\,(1-\frac{1}{3})\,\sqrt{8\,r_{0}}}\ =\ \frac{%
3^{-k}}{2\,\sqrt{8\,r_{0}}}.  \label{eq-4.34}
\end{equation}%
We hence conclude that 
\begin{equation}
\sup_{x\in \omega ({\underline{r}})}\bigg\{\sum_{y\in \omega ({\underline{r}}%
)\setminus \{x\}}\big|\big\la\delta _{x}\,\big|\;\big(\rho +h(\mathfrak{e})%
\big)^{-1}\delta _{y}\big\ra\big|\bigg\}\bigg)\ \leq \ \frac{C_{\ref{lem-4.3}%
}(d,\mathfrak{e})}{2\,\sqrt{8\,r_{0}}}.  \label{eq-4.35}
\end{equation}%
Thus, the operator norm of the Birman-Schwinger operator is strictly smaller
than one, 
\begin{equation}
\big\|V^{1/2}\,(\rho +h(\mathfrak{e}))^{-1}\,V^{1/2}\big\|\ \leq \
|V|_{\infty }\,\bigg(\frac{1}{\eta (\mathfrak{e})}\;+\;\frac{C_{\ref{lem-4.3}%
}(d,\mathfrak{e})}{2\,\sqrt{8\,r_{0}}}\bigg)\ <\ 1,  \label{eq-4.36}
\end{equation}%
for all $\rho >0$, which implies that $N[\mathfrak{e},V]=0$. \hfill $\square 
$\bigskip

The last lemma has the following immediate consequences.

\begin{corollary}[Thm.~\protect\ref{thm-1.9}]
\label{cor-4.5} Let $d\geq 3$ and $\mathfrak{e}$ be an admissible
dispersion. Then there exists a potential $V\notin \bigcup\limits_{p\geq
1}\ell^{p}(\Gamma )$ with $N[\mathfrak{e},V]=0$.
\end{corollary}

\noindent \textit{Proof:} Fix $r_{0}\in \mathbb{N}$, choose $r_{k}\doteq
9^{k}\,r_{0}$, $x_{k}\doteq (r_{k},0,\ldots ,0)$, and set 
\begin{equation}
V(x)\ \doteq \sum_{j=0}^{\infty }\mathbf{1}_{\{x_{j}\}}(x)\,\frac{\eta (%
\mathfrak{e})}{\ln (4+j)}.  \label{eq-4.37}
\end{equation}%
Note that $V\in \ell _{0}^{\infty }(\Gamma )$ but that, for all $p\geq 1$,
the $p$-norm of $V$ diverges, $|V|_{p}=\eta (\mathfrak{e})^{p}\sum_{j=0}^{%
\infty }\big[\ln (4+j)\big]^{-p}=\infty $. Moreover, $|V|_{\infty }=\frac{1}{%
\ln (4)}\eta (\mathfrak{e})<\eta (\mathfrak{e})$, and Lemma~\ref{lem-4.4}
implies that $N[\mathfrak{e},V]=0$ provided $r_{0}\in \mathbb{N}$ is chosen
sufficiently large such that $C_{\ref{lem-4.3}}(d,\mathfrak{e})\,\eta (%
\mathfrak{e})r_{0}^{-1/2}<4\left( 1-\frac{1}{\ln (4)}\right) $. \hfill $%
\square $\bigskip

\noindent We remark that $N_{sc}[\mathfrak{e},V]=\infty $ in Corollary~\ref%
{cor-4.5}, since $V\notin \bigcup\limits_{p\geq 1}\ell ^{p}(\Gamma )$. Thus,
a \textit{lower} bound on $N[\mathfrak{e},V]$ in terms of $\ell ^{p}$-norms
or in multiples of $N_{sc}[\mathfrak{e},V]$ cannot, in general, hold true.
See also \cite[Eq. (1.8)]{RozenblumSolomyak2009}.

\begin{corollary}[Thm.~\protect\ref{thm-1.5}]
\label{cor-4.5,1} Let $d\geq 3$, $\mathfrak{e}$ be an admissible dispersion.
Given ${\varepsilon }\in (0,1)$ and a potential $V\in \ell _{0}^{\infty
}(\Gamma ,\mathbb{R}_{0}^{+})$, there exists a rearrangement ${\widetilde{V}}%
\in \ell _{0}^{\infty }(\Gamma ,\mathbb{R}_{0}^{+})$ of $V$ such that 
\begin{equation}
N[\mathfrak{e},{\widetilde{V}}]\ \leq \ \mathcal{L}_{\widetilde{V}}[(1-{%
\varepsilon })\eta (\mathfrak{e})]\ =\ \#\big\{x\in \Gamma \,\big|\ V(x)\geq
(1-{\varepsilon })\eta (\mathfrak{e})\big\}.  \label{eq-4.37,1}
\end{equation}
\end{corollary}

\noindent \textit{Proof:} We write $V=V^{(>)}+V^{(<)}$ with 
\begin{equation}
V^{(>)}\ =\ V\cdot \mathbf{1}\big[V\geq (1-{\varepsilon })\eta (\mathfrak{e})%
\big]\quad \text{and}\quad V^{(<)}\ =\ V\cdot \mathbf{1}\big[V<(1-{%
\varepsilon })\eta (\mathfrak{e})\big].  \label{eq-4.37,2}
\end{equation}%
Note that $V^{(>)}$ has bounded support. Thus, choosing ${\widetilde{V}}%
^{(<)}$ to be a rearrangement of $V^{(<)}$ with 
\begin{equation}
\mathop{\mathrm{supp}} {\widetilde{V}}^{(<)}\ \subset \ \big\{%
(r_{k},0,\ldots ,0)\;\big|\ r_{k}\doteq 9^{k}\,r_{0}\,,\ k\in \mathbb{N}_{0}%
\big\}  \label{eq-4.37,3}
\end{equation}%
and $r_{0}\in \mathbb{N}$ chosen sufficiently large, we find that 
\begin{equation}
\big\|{\widetilde{V}}^{(<)}\big\|_{\infty }\ =\ (1-{\varepsilon })\eta (%
\mathfrak{e})\ <\ \eta (\mathfrak{e})-\frac{1}{4}\,C_{\ref{lem-4.3}}(d,%
\mathfrak{e})\,\eta (\mathfrak{e})^{2}r_{0}^{-1/2},  \label{eq-4.37,4}
\end{equation}%
and Lemma~\ref{lem-4.4} implies that $N[\mathfrak{e},{\widetilde{V}}%
^{(<)}]=0 $. Hence, defining ${\widetilde{V}}\doteq V^{(>)}+{\widetilde{V}}%
^{(<)}$, we have for sufficiently large $r_{0}\in \mathbb{N}$ that $\mathrm{%
supp\;}V^{(>)}\cap \mathrm{supp\;}{\widetilde{V}}^{(<)}=\varnothing $, ${%
\widetilde{V}}$ is a rearrangement of $V$, and 
\begin{eqnarray}
N[\mathfrak{e},{\widetilde{V}}]\ &\leq &\ \#\mathop{\mathrm{supp}}%
V^{(>)}\,+\,N[\mathfrak{e},{\widetilde{V}}^{(<)}]\   \label{eq-4.37,5} \\
&=&\ \#\mathop{\mathrm{supp}}V^{(>)}\ =\ \mathcal{L}_{V}[(1-{\varepsilon }%
)\eta (\mathfrak{e})],
\end{eqnarray}%
by Corollary~\ref{cor-3.2}. \hfill $\square $\bigskip

\noindent The next theorem illustrates for $d\geq 3$ that -- opposed to the
continuum case -- the asymptotics of $N[\mathfrak{e},\lambda V]$ as $\lambda
\rightarrow \infty $ can be prescribed arbitrarily.

\begin{theorem}[Thm.~\protect\ref{thm-1.10}]
\label{thm-4.6} Let $d\geq 3$ and $\mathfrak{e}$ be any admissible
dispersion. Let further $F:[1,\infty )\rightarrow \mathbb{N}$ be an
arbitrary monotonically increasing, positively integer-valued,
right-continuous function. Then, for any $\varepsilon \in (0,1/2)$, there
exists a potential $V_{F,\varepsilon }\in \ell _{0}^{\infty }(\Gamma ,%
\mathbb{R}
_{0}^{+})$ such that 
\begin{equation}
\forall \, {\lambda \geq 2}:\;F\big((1-\varepsilon )\lambda \big)\ \leq \ N%
\big[\mathfrak{e},\lambda V_{F}\big]\ \leq \ F\big((1+\varepsilon )\lambda %
\big).  \label{eq-4.38}
\end{equation}
\end{theorem}

\noindent \textit{Proof:} For the proof, we abbreviate $\eta \doteq \eta (%
\mathfrak{e})$. Since $F:[1,\infty )\rightarrow \mathbb{N}$ is monotonically
increasing and right-continuous, there is a monotonically increasing
sequence $1\leq \lambda _{1}\leq \lambda _{2}\leq \lambda _{3}\leq \cdots $
such that 
\begin{equation}
F(\lambda )\ =\ \sum_{j=1}^{\infty }\mathbf{1}[\lambda _{j}\leq \lambda ].
\label{eq-4.39}
\end{equation}%
Note that the monotonicity of $F$ is not necessarily strict, and possibly $%
\lambda _{j}=\lambda _{j+1}$. For a sequence ${\underline{r}}%
=(r_{k})_{k=0}^{\infty }$ of positive integers, with $9r_{k}\leq r_{k+1}$,
to be further specified later, and $x_{k}=(r_{k},0,\ldots ,0)\in \Gamma $,
we set 
\begin{equation}
V_{F,\varepsilon }(x)\ \doteq \ \sum_{j=1}^{\infty }\frac{\eta }{\lambda _{j}%
}\mathbf{1}_{\{x_{j}\}}(x).  \label{eq-4.40}
\end{equation}%
Let $\varepsilon^{\prime }>0$ be such that $(1+\varepsilon^{\prime
})^{-1}>1-\varepsilon $. Choosing $r_{0}>0$ large enough such that%
\begin{equation*}
\eta \,\sup_{\rho >0}\sup_{x\in \mathrm{supp\;}V_{F,\varepsilon }}\sum_{y\in 
\mathrm{supp\;}V_{F,\varepsilon }\backslash \{x\}}|\langle \delta
_{x}|\,(\rho +h(\mathfrak{e}))^{-1}\delta _{y}\rangle |<\frac{\varepsilon
^{\prime }}{1+\varepsilon^{\prime }}
\end{equation*}%
we observe that 
\begin{eqnarray}
\mathcal{L}_{\lambda V_{F,\varepsilon }}((1+\varepsilon^{\prime })\eta ) &=&%
\mathcal{L}_{V_{F,\varepsilon }}((1+\varepsilon^{\prime })\eta /\lambda )\ \;
\\
&=&\ \;\#\bigg\{x\in \Gamma \bigg|\ V_{F,\varepsilon }(x)\geq (1+\varepsilon
^{\prime })\frac{\eta }{\lambda }\bigg\} \\
&=&\sum_{j=1}^{\infty }\mathbf{1}\bigg[\frac{\eta }{\lambda _{j}}\geq
(1+\varepsilon^{\prime })\frac{\eta }{\lambda }\bigg]\ \;=\
\;F((1+\varepsilon^{\prime })^{-1}\lambda ).  \notag
\end{eqnarray}%
Thanks to Lemma~\ref{lem-3.6.b}, we have thus established the lower bound on 
$N[\mathfrak{e},\lambda V_{F,\varepsilon }]$ in \eqref{eq-4.38}, 
\begin{equation}
F((1-\varepsilon )\lambda )\ \leq F((1+\varepsilon^{\prime })^{-1}\lambda )\
\leq \ N\big[\mathfrak{e},\lambda V_{F,\varepsilon }\big]  \label{eq-4.42}
\end{equation}%
for all $\lambda \geq 2$. Choose now $\varepsilon^{\prime }>0$ such that $%
(1-\varepsilon^{\prime })^{-1}<1+\varepsilon $. For the proof of the upper
bound in \eqref{eq-4.38} we write $\lambda V_{F,\varepsilon }=V_{F,\lambda
}^{(>)}+V_{F,\lambda }^{(<)}$, where 
\begin{eqnarray}
V_{F,\lambda }^{(>)}(x) &\doteq &\lambda \,V_{F,\varepsilon }\,\mathbf{1}%
\Big[V_{F,\varepsilon }(x)\geq (1-\varepsilon^{\prime })\frac{\eta }{\lambda 
}\Big]  \label{eq165} \\[0.02in]
&=&\sum_{j=1}^{\infty }\mathbf{1}_{\{x_{j}\}}(x)\,\mathbf{1}\big[\lambda
_{j}\leq (1-\varepsilon^{\prime })^{-1}\lambda \big]\frac{\eta \,\lambda }{%
\lambda _{j}},  \notag \\
V_{F,\lambda }^{(<)}(x) &\doteq &\lambda \,V_{F,\varepsilon }\,\mathbf{1}%
\Big[V_{F,\varepsilon }<(1-\varepsilon^{\prime })\frac{\eta }{\lambda }\Big]
\notag \\[1ex]
&=&\sum_{j=1}^{\infty }\mathbf{1}_{\{x_{j}\}}(x)\,\mathbf{1}\big[\lambda
_{j}>(1-\varepsilon^{\prime })^{-1}\lambda \big]\frac{\eta \,\lambda }{%
\lambda _{j}},  \notag
\end{eqnarray}%
Observe that, due to (\ref{eq165}) 
\begin{equation}
\#\mathop{\mathrm{supp}}V_{F,\lambda }^{(>)}\ =\ \#\bigg\{x\in \omega ({%
\underline{r}})\bigg|\ V_{F,\varepsilon }(x)\geq (1-\varepsilon^{\prime })%
\frac{\eta }{\lambda }\bigg\}\ =\ F\big((1-\varepsilon^{\prime
})^{-1}\lambda \big).  \label{eq-4.45}
\end{equation}%
Hence, Corollary~\ref{cor-3.2} yields 
\begin{equation}
N\big[\mathfrak{e},\lambda V_{F,\varepsilon }\big]\ \leq \ F\big(%
(1+\varepsilon )\lambda \big)+N\big[\mathfrak{e},V_{F,\lambda }^{(<)}\big],
\label{eq-4.46}
\end{equation}%
and it remains to fix the sequence ${\underline{r}}$ so that 
\begin{equation}
N\big[\mathfrak{e},V_{F,\lambda }^{(<)}\big]\ =\ 0,  \label{eq-4.47}
\end{equation}%
for all $\lambda \geq 1$. To this end, we first note that 
\begin{equation}
\big\|V_{F,\lambda }^{(<)}\big\|_{\infty }\ \leq \ \eta \,(1-\varepsilon
^{\prime }).  \label{eq-4.48}
\end{equation}%
From Lemma~\ref{lem-4.4}, (\ref{eq-4.47}) holds by choosing $r_{0}>0$ large
enough and the right-hand inequality in \eqref{eq-4.38} follows. \hfill $%
\square $\bigskip

A similar result in proven in \cite[Section 6]{RozenblumSolomyak2009}.
Observe, however, that, in contrast to \cite{RozenblumSolomyak2009}, we do
not assume that $\lambda _{j}/\lambda _{j+1}\rightarrow 1$, as $j\rightarrow
\infty $, for the asymptotics of eigenvalues. Moreover, the positivity
preserving property of the hopping matrix $h(\mathfrak{e})$ is not needed.

Assume that for a given potential $V\in \ell _{0}^{\infty }(\Gamma ,%
\mathbb{R}
_{0}^{+})$, $N[\mathfrak{e},\lambda V]\sim N_{sc}[\mathfrak{e},\lambda
V]<\infty $ at large $\lambda >0$, i.e., that $N[\mathfrak{e},\lambda V]$ is
finite and obeys the Weyl asymptotics at large $\lambda $. Then it would
follow that $N[\mathfrak{e},\lambda V]=\mathcal{O}(\lambda ^{d/2})$. By the
last theorem, for any $\alpha >0$, there are potentials $V_{\alpha }\in \ell
_{0}^{\infty }(\Gamma ,%
\mathbb{R}
_{0}^{+})$ such that $N[\mathfrak{e},\lambda V]$ behaves like $\lambda
^{\alpha }$ as $\lambda \rightarrow \infty $. In particular, the
semi-classical asymptotics cannot hold for $V_{\alpha }$ with $\alpha >d/2$.
See also \cite[Eq. (1.8)]{RozenblumSolomyak2009}. Observe, however, that in
such a case, by the semi-classical upper bound on $N[\mathfrak{e},\lambda
V_{\alpha }]$ (Theorem~\ref{thm-1.1}), $N_{sc}[\mathfrak{e},\lambda
V_{\alpha }]=\infty $ (whereas $N[\mathfrak{e},\lambda V_{\alpha }]<\infty $%
) for all $\lambda >0$ and speaking about semi-classical behavior does not
really make sense. We discuss below another kind of example for which the
semi-classical asymptotics -- in the sense of two--side bounds -- is
violated, even if $N_{sc}[\mathfrak{e},\lambda V]<\infty $ for all $\lambda
>0$.

\begin{theorem}[Thm. \protect\ref{thm-liminf-limsup}]
\label{thm-non-sc} Let $d\geq 3$ and $\mathfrak{e}$ be any admissible
dispersion relation. There is a potential $V\geq 0$, $V\in \ell
^{d/2}(\Gamma ,%
\mathbb{R}
_{0}^{+})$, such that 
\begin{equation}
\liminf_{\lambda \rightarrow \infty }\frac{N_{sc}[\mathfrak{e},\lambda V]}{N[%
\mathfrak{e},\lambda V]}<\infty ,\quad \limsup_{\lambda \rightarrow \infty }%
\frac{N_{sc}[\mathfrak{e},\lambda V]}{N[\mathfrak{e},\lambda V]}=\infty .
\end{equation}
\end{theorem}

\noindent \textit{Proof:} Define the potentials $V_{1},V_{2}\in \ell
^{d/2}(\Gamma ,%
\mathbb{R}
_{0}^{+})$ by 
\begin{equation*}
V_{1}(x)\doteq \frac{1}{\left\langle x\right\rangle^{2}\ln \left\langle
x\right\rangle },\;V_{2}(x)\doteq \mathrm{e}^{-|x|}.
\end{equation*}%
Clearly, $g_{-}(V_{1})=1$ and $g_{+}(V_{2})=0$. By Lemma~\ref{lem-4.1},%
\begin{equation}
\lim_{\lambda \rightarrow \infty }\frac{N_{sc}^{>}[\mathfrak{e},\lambda
V_{1}]}{N_{sc}[\mathfrak{e},\lambda V_{1}]}=0,\;\lim_{\lambda \rightarrow
\infty }\frac{N_{sc}^{>}[\mathfrak{e},\lambda V_{2}]}{N_{sc}[\mathfrak{e}%
,\lambda V_{2}]}>0.  \label{bound lim V1V2}
\end{equation}%
For any monotonically increasing sequence $\alpha =(\alpha _{n})_{n\in 
\mathbb{N}}$ of positive real numbers define $\beta _{\alpha }:\Gamma
\rightarrow \{0,1\}$ by $\beta _{\alpha }(x)\doteq 1$ if $\alpha _{1+2n}\leq
|x|\leq \alpha _{2+2n}$ for some $n\in \mathbb{N}_{0}$, and $\beta _{\alpha
}(x)\doteq 0$ else. Consider potentials of the form $\tilde{V}=\tilde{V}%
_{\alpha }\doteq \beta _{\alpha }(V_{1}-V_{2})+V_{2}\geq 0$. By (\ref{bound
lim V1V2}), there exists a sequence $\alpha $ such that: 
\begin{equation}
\liminf_{\lambda \rightarrow \infty }\frac{N_{sc}[\mathfrak{e},\lambda 
\tilde{V}]}{N_{sc}^{>}[\mathfrak{e},\lambda \tilde{V}]}<\infty
,\;\limsup_{\lambda \rightarrow \infty }\frac{N_{sc}[\mathfrak{e},\lambda 
\tilde{V}]}{N_{sc}^{>}[\mathfrak{e},\lambda \tilde{V}]}=\infty .
\label{eq173}
\end{equation}

By (\ref{eq173}) and Lemma~\ref{lem-3.6}, for any rearrangement $V$ of $%
\tilde{V}$,%
\begin{equation*}
\liminf_{\lambda \rightarrow \infty }\frac{N_{sc}[\mathfrak{e},\lambda V]}{N[%
\mathfrak{e},\lambda V]}<\infty .
\end{equation*}%
Observe that, by Corollary~\ref{cor-3.2} and Lemma~\ref{lem-4.4}, there is a
rearrangement $V$ of $\tilde{V}$ such that 
\begin{equation}
\limsup\limits_{\lambda \rightarrow \infty }\frac{N_{sc}^{>}[\mathfrak{e}%
,\lambda (2\mathfrak{e}_{\mathrm{\max }}/\eta (\mathfrak{e}))V]}{N[\mathfrak{%
e},\lambda V]}\geq 1.  \label{eq173b}
\end{equation}%
To conclude the proof use that for some $1<C<\infty $, 
\begin{equation}
C^{-1}N_{sc}[\mathfrak{e},\lambda V]\leq N_{sc}[\mathfrak{e},\lambda (2%
\mathfrak{e}_{\mathrm{\max }}/\eta (\mathfrak{e}))V]\leq C\,N_{sc}[\mathfrak{%
e},\lambda V]
\end{equation}%
for all $\lambda >0$. This together with (\ref{eq173}) and (\ref{eq173b})
imply%
\begin{equation*}
\limsup_{\lambda \rightarrow \infty }\frac{N_{sc}[\mathfrak{e},\lambda V]}{N[%
\mathfrak{e},\lambda V]}=\infty .
\end{equation*}%
Note that we have used above the invariance of the semi-classical quantities 
$N_{sc}^{>}[\mathfrak{e},\tilde{V}]$ and $N_{sc}[\mathfrak{e},\tilde{V}]$
w.r.t. rearrangements of $\tilde{V}$. \hfill $\square $

\section{One and Two Dimensions}

\label{sec-5}

We start this section by showing (Corollary~\ref{cor-5.3}) that the
semi-classical upper bound, as stated in Theorem~\ref{thm-1.1} for instance,
cannot be valid in less than three dimensions.

\begin{lemma}
\label{lem-5.1} Let $d\in \{1,2\}$, $\mathfrak{e}$ be an admissible
dispersion relation, and $V\geq 0$ be a potential with finite support. For
all $\rho >0$ and all rearrangements ${\widetilde{V}}$ of $V$ define the
compact self-adjoint operator 
\begin{equation}
K(\rho ,{\widetilde{V}})\ =\ P_{\mathop{\mathrm{Ran}} {\widetilde{V}}}\,{%
\widetilde{V}}^{1/2}\,(\rho +h(\mathfrak{e}))^{-1}\,{\widetilde{V}}%
^{1/2}\,P_{\mathop{\mathrm{Ran}} {\widetilde{V}}}\;-\;P_{\mathop{%
\mathrm{Ran}} {\widetilde{V}}}.  \label{eq-5.1}
\end{equation}%
Then there exist $\rho >0$ and a rearrangement ${\widetilde{V}}$ of $V$ such
that $K(\rho ,{\widetilde{V}})>0$.
\end{lemma}

\noindent \textit{Proof}: If $\mathop{\mathrm{supp}}V=\varnothing $ there is
nothing to prove, so we assume that $V\neq 0$. Let ${\widetilde{V}}\geq 0$
be a rearrangement of $V$. Then for all $\rho >0$ and all $\psi =(\psi
_{x})_{x\in \Gamma }\in \mathop{\mathrm{Ran}} {\widetilde{V}}$, 
\begin{eqnarray}
\langle \psi \,|\,K(\rho ,{\widetilde{V}})\,\psi \rangle &=&-|\psi
|_{2}^{2}\,+\,\sum_{x\in \mathop{\mathrm{supp}} {\widetilde{V}}}{\widetilde{V%
}}(x)|\psi _{x}|^{2}\int_{\Gamma^{\ast }}\frac{\mathrm{d}\mu^{\ast }(p)}{%
\rho +\mathfrak{e}(p)}  \label{eq-5.2} \\
&&+\sum\limits_{x,y\in \mathop{\mathrm{supp}} {\widetilde{V}},\;x\neq y}[{%
\widetilde{V}}(x){\widetilde{V}}(y)]^{1/2}\big\la\delta _{x}\big|(\rho +h(%
\mathfrak{e}))^{-1}\delta _{y}\big\ra\overline{\psi _{x}}\psi _{y},  \notag
\end{eqnarray}%
and thus 
\begin{eqnarray}
K(\rho ,{\widetilde{V}}) &\geq &-1+\min_{x\in \mathop{\mathrm{supp}}%
V}V(x)\,\int_{\Gamma^{\ast }}\frac{\mathrm{d}\mu^{\ast }(p)}{\rho +\mathfrak{%
e}(p)}  \label{5.3} \\
&&-|V|_{\infty }\sup\limits_{\psi \in \mathop{\mathrm{Ran}} {\widetilde{V}}%
,\;|\psi |_{2}=1}\sum\limits_{x,y\in \mathop{\mathrm{supp}}\widetilde{V}%
,\;x\neq y}\big|\big\la\delta _{x}\big|(\rho +h(\mathfrak{e}))^{-1}\delta
_{y}\big\ra\overline{\psi _{x}}\psi _{y}\big|.  \notag
\end{eqnarray}%
Choose $\rho >0$ such that 
\begin{equation}
\min_{x\in \mathop{\mathrm{supp}}V}V(x)\int_{\Gamma^{\ast }}\frac{\mathrm{d}%
\mu^{\ast }(p)}{\rho +\mathfrak{e}(p)}\ >\ 2.  \label{eq-5.4}
\end{equation}%
This is always possible since $d\leq 2$. For any fixed $\rho >0$, we have
that 
\begin{equation*}
\langle \delta _{x}|(\rho +h(\mathfrak{e}))^{-1}\delta _{y}\rangle
\rightarrow 0
\end{equation*}%
as $|x-y|\rightarrow \infty $. This follows from the Riemann-Lebesgue Lemma
since $\langle \delta _{x}|(\rho +h(\mathfrak{e}))^{-1}\delta _{y}\rangle $
is the Fourier transform of the integrable function $(\rho +\mathfrak{e}%
)^{-1}\in L^{1}(\Gamma^{\ast })$. In particular, there is a rearrangement ${%
\widetilde{V}}$ of $V$ such that 
\begin{equation}
|V|_{\infty }\sup\limits_{\psi \in \mathop{\mathrm{Ran}} {\widetilde{V}}%
,\;|\psi |_{2}=1}\sum\limits_{x,y\in \mathop{\mathrm{supp}}\widetilde{V}%
,\;x\neq y}\big|\big\la\delta _{x}\big|(\rho +h(\mathfrak{e}))^{-1}\delta
_{y}\big\ra\overline{\psi _{x}}\psi _{y}\big|\ \leq \ 1.  \label{eq-5.5}
\end{equation}%
For such $\rho >0$ and ${\widetilde{V}}$ we hence have that $K(\rho ,{%
\widetilde{V}})>0$. \hfill $\square $

\begin{theorem}
\label{thm-5.2} Let $d\in \{1,2\}$ and $\mathfrak{e}$ be an admissible
dispersion relation. Then, for any finitely supported potential $V$, there
is a rearrangement ${\widetilde{V}}$ of $V$ such that 
\begin{equation}
N[\mathfrak{e},{\widetilde{V}}]\ =\ \#\mathop{\mathrm{supp}} {\widetilde{V}}%
\ =\ \#\mathop{\mathrm{supp}}V.  \label{eq-5.5,1}
\end{equation}
\end{theorem}

\noindent \textit{Proof}: Clearly, for any rearrangement ${\widetilde{V}}$
of $V$, we have $N[\mathfrak{e},{\widetilde{V}}]\leq \#\mathop{\mathrm{supp}}%
V$, as follows, e.g., from Corollary~\ref{cor-3.2} and the fact that $N[%
\mathfrak{e},0]=0$. Let $\rho >0$ and the rearrangement ${\widetilde{V}}$ of 
$V$ be as in the lemma above. Then, by the min-max principle and the bound $%
K(\rho ,{\widetilde{V}})>0$, the compact operator $({\widetilde{V}}%
)^{1/2}(\rho +h(\mathfrak{e}))^{-1}({\widetilde{V}})^{1/2}$ has at least $%
\mathrm{dim}\;\mathop{\mathrm{Ran}}{\widetilde{V}}=|\mathop{\mathrm{supp}}\,{%
\widetilde{V}}|$ discrete eigenvalues above $1$. By Lemma~\ref{lem-2.2}, it
follows from this that $N[\mathfrak{e},{\widetilde{V}}]\geq \#%
\mathop{\mathrm{supp}}{\widetilde{V}}$. \hfill $\square $\bigskip

Observing that the semi-classical number of bound states $N_{sc}[\mathfrak{e}%
,V]$ is invariant w.r.t. rearrangements of the potential $V$, the following
corollary follows immediately:

\begin{corollary}[Breakdown of the semi-classical upper bound in $d=1,2$]
\label{cor-5.3} Let $d\in \{1,2\}$ and $\mathfrak{e}$ be any admissible
dispersion. Then, for all $\epsilon >0$, 
\begin{equation}
\sup \bigg\{\frac{N[\mathfrak{e},V]}{N_{sc}[\mathfrak{e},V]}\;\bigg|\
V,\;N_{sc}[\mathfrak{e},V]<\epsilon \bigg\}\ =\ \infty .  \label{eq-5.6}
\end{equation}
\end{corollary}

The last corollary implies in one or two dimensions that multiples of $%
N_{sc}[\mathfrak{e},V]$ cannot be, in general, an upper bound on $N[%
\mathfrak{e},V]$. The discussion above shows, more precisely, that $\mathrm{%
const\,}N_{sc}[\mathfrak{e},V]$ fails to be such an upper bound in the case
of sparse potentials, i.e. in the situation where the distance between
points in the support of the potential $V$ is large. Hence, any quantity $%
Q(V)$ which is supposed to be an upper bound on $N[\mathfrak{e},V]$ should
keep track of the behavior of $V$ in space. This motivates the use of the
weighted semi-classical quantities $N_{sc}\big[\mathfrak{e},\tilde{V}(V)%
\big]
$ -- as stated in Theorem~\ref{thm-1.2} -- as upper bounds on $N[\mathfrak{e}%
,V]$ in one and two dimensions.

For any $p>0$, $m\geq 0$, and any function $V:\Gamma \rightarrow \mathbb{R}%
_{0}^{+}$ define 
\begin{equation}
|V|_{p,m}\ \doteq \ \bigg(\sum_{x\in \Gamma }V^{p}(x)\,\langle x\rangle^{m}%
\bigg)^{1/p}.  \label{eq-5.7}
\end{equation}%
Observe that $|\cdot |_{p,m}$ is not a norm, for $p\in (0,1)$, but only a
homogeneous functional of degree one. For any function $\mathfrak{e}\in
C^{m}(\Gamma^{\ast },{\mathbb{C}})$ and $m\in {\mathbb{N}}_{0}$, define the $%
C^{m}$-(semi)norms by 
\begin{equation}
\Vert \mathfrak{e}\Vert _{C^{m}}\ \doteq \ \max\limits_{\underline{n}\in {%
\mathbb{N}}_{0}^{d},\;|\underline{n}|=m}\;\max_{p\in \Gamma^{\ast }}\big|%
\partial _{p}^{\underline{n}}\mathfrak{e}(p)\big|.  \label{eq-5.8}
\end{equation}

Let $\mathfrak{e}$ be an admissible dispersion relation. We denote the set
of all critical points of $\mathfrak{e}$ by 
\begin{equation}
\mathrm{Crit}(\mathfrak{e})\ \doteq \ \big\{p\in \Gamma^{\ast }\;\big|\
\nabla \mathfrak{e}(p)=0\big\}.  \label{eq-5.9}
\end{equation}%
Recall that, as $\Gamma^{\ast }$ is compact, dispersion relations have at
most finitely many critical points. $\mathrm{Min}(\mathfrak{e})\subset 
\mathrm{Crit}(\mathfrak{e})$ denotes the set of points on which the minimum
of $\mathfrak{e}$ is taken.

Let $\mathfrak{e}^{\prime \prime }(p)$ be the Hessian matrix of $\mathfrak{e}
$ at $p\in \text{Crit}(\mathfrak{e})$. Define the \textit{minimal curvature
of\ (the graph of) $\mathfrak{e}$ at $p\in \mathrm{Crit}(\mathfrak{e})$} by 
\begin{equation}
K(\mathfrak{e},p)\ \doteq \ \min \big\{|\lambda |^{1/2}\;\big|\ \lambda \in
\sigma \lbrack \mathfrak{e}^{\prime \prime }(p)]\big\}>0.  \label{eq-5.10}
\end{equation}%
Define also the \textit{minimal (critical) curvature of $\mathfrak{e}$} by 
\begin{equation}
K(\mathfrak{e})\ \doteq \ \min \big\{K(\mathfrak{e},p)\;\big|\ p\in \mathrm{%
Crit}(\mathfrak{e})\big\}>0.  \label{eq-5.10,1}
\end{equation}

\begin{lemma}[A priori upper bound on $N{(e,V)}$, $d=1,2$]
\label{lem-5.4} Let $\mathfrak{e}$ be any dispersion relation from $%
C^{3}(\Gamma ^{\ast },{\mathbb{R}})$. Let $C<\infty $ and $K>0$ be such that 
$\Vert \mathfrak{e}\Vert _{C^{3}}<C$ and $K(\mathfrak{e})>K$. Define $\delta
\doteq \min \{\mathfrak{e}(p)\;|\;p\in \mathrm{Crit}(\mathfrak{e})\backslash 
\mathrm{Min}(\mathfrak{e})\}>0$.

\begin{enumerate}
\item[(i)] There is a constant $C_{\ref{lem-5.4}(i)}<\infty $ depending only
on $\mathfrak{e},C,K,\#\mathrm{Min}(\mathfrak{e})$, and $\delta $ such that $%
N[\mathfrak{e},V]\leq \#\mathrm{Min}(\mathfrak{e})$ whenever $|V|_{1/2,1}<C_{%
\ref{lem-5.4}a}$.

\item[(ii)] There is a constant $C_{\ref{lem-5.4}(ii)}<\infty $ depending
only on $\mathfrak{e},C,K,\#\mathrm{Min}(\mathfrak{e})$, and $\delta $ such
that 
\begin{equation}
N[\mathfrak{e},V]\ \leq \ C_{\ref{lem-5.4}(ii)}\,|V|_{1/2,2}\,+\,\#\mathrm{%
Min}(\mathfrak{e}).  \label{eq-5.11}
\end{equation}
\end{enumerate}
\end{lemma}

\noindent \textit{Proof:}\newline
Let $C^{1}(\Gamma^{\ast })$ be the Banach space of all continuously
differentiable functions $\Gamma^{\ast }\rightarrow {\mathbb{C}}$ with norm $%
\Vert \cdot \Vert _{C_{1}}$. Observe that if $|V|_{1/2,1}$ is finite $%
\mathcal{F}^{\ast }\circ V^{1/2}$ defines a continuous linear map $\ell
^{2}(\Gamma )\rightarrow C_{1}(\Gamma^{\ast })$ with 
\begin{equation}
\Vert \mathcal{F}^{\ast }\circ V^{1/2}\Vert _{\mathcal{B}[\ell^{2}(\Gamma
),C_{1}(\Gamma^{\ast })]}\leq |V|_{1/2,1}^{1/2}.  \label{V.C1}
\end{equation}%
Let $\mathrm{Min}(\mathfrak{e})=$ $\{p^{(1)},\ldots ,p^{(m)}\}$, $m=\,\#%
\mathrm{Min}(\mathfrak{e})$, and define the linear functionals $\zeta _{i}$, 
$i=1,2,$ $\ldots ,m$, on $\ell^{2}(\Gamma )$ by $\zeta _{i}({\varphi }%
)\doteq \mathcal{F}^{\ast }\circ V^{1/2}({\varphi })(p^{(i)})$. By (\ref%
{V.C1}), the functionals $\zeta _{i}$ are continuous. Let $%
X=\bigcap_{i=1}^{m}\mathrm{\ker }\,\zeta _{i}$. Assume that $H(\mathfrak{e}%
,V)$ has more than $m$ eigenvalues (counting multiplicities) below $0$.
Then, by Lemma~\ref{lem-2.2} and the min-max principle, there is some $\rho
>0$ and some $(m+1)$--dimensional subspace $S\subset \ell^{2}(\Gamma )$ with 
\begin{equation}
\min\limits_{{\varphi }\in S,\;|{\varphi }|_{2}=1}\langle {\varphi }%
\,|\,V^{1/2}(\rho +h(\mathfrak{e}))^{-1}V^{1/2}{\varphi }\rangle >1.
\end{equation}%
Observe that for all ${\varphi }\in \ell^{2}(\Gamma )$, 
\begin{equation}
\langle {\varphi }\,|\,V^{1/2}(\rho +h(\mathfrak{e}))^{-1}V^{1/2}{\varphi }%
\rangle =\int_{\Gamma^{\ast }}\frac{|\mathcal{F}^{\ast }\circ V^{1/2}({%
\varphi })(p)|^{2}}{\rho +\mathfrak{e}(p)}\mathrm{d}\mu^{\ast }(p).
\end{equation}%
As the dimension of $S$ is larger than $m$, there is a vector $\tilde{{%
\varphi }}\in S\cap X$, $|\tilde{{\varphi }}|_{2}=1$. Notice that in this
case there is a constant $\mathrm{const}<\infty $ depending only on $C$ and $%
m$ such that for all $p\in \Gamma^{\ast }$, 
\begin{equation}
|\mathcal{F}^{\ast }\circ V^{1/2}(\tilde{{\varphi }})(p)|^{2}\leq \mathrm{%
const}\,|V|_{1/2,1}\prod_{i=1}^{m}(1-\cos (p-p^{(i)})),
\end{equation}%
where for each $q=(q_{1},\ldots ,q_{d})\in \Gamma _{d}^{\ast }$, 
\begin{equation*}
\cos (q)\doteq d^{-1}(\cos (q_{1})+\ldots +\cos (q_{d})).
\end{equation*}%
It means that 
\begin{equation}
1<\mathrm{const}\,|V|_{1/2,1}\int_{\Gamma^{\ast }}\frac{\prod_{i=1}^{m}(1-%
\cos (p-p^{(i)}))}{\rho +\mathfrak{e}(p)}\mathrm{d}\mu^{\ast }(p).
\label{bound.int.d12}
\end{equation}%
Observing that the integral on the right-hand side of (\ref{bound.int.d12})
is bounded by a constant depending only on $C,K$ and $m$ this concludes the
proof of (i).

Now we prove (ii). For any $q\in \Gamma^{\ast }$ define the linear maps $%
\zeta _{q}^{\prime}: \ell^2(\Gamma )\rightarrow {\mathbb{C}}\times {\mathbb{C%
}}^{d} $ by 
\begin{equation}
\zeta _{q}^{\prime }({\varphi })=\Big((\mathcal{F}^{\ast }\circ V^{1/2})({%
\varphi })(q)\,,\,(\nabla \mathcal{F}^{\ast }\circ V^{1/2})({\varphi })(q)%
\Big).
\end{equation}%
By $|V|_{1/2,1}\leq |V|_{1/2,2}<\infty $ it follows that $\zeta _{q}^{\prime
}$ is continuous.

There is a constant $\mathrm{const}<\infty $ such that, for any fixed $\mu
>0 $ small enough, there is a set of points $\{q_{1},\ldots ,q_{n(\mu )}\}$
from $\Gamma^{\ast }$ containing $\mathrm{Min}(\mathfrak{e})$ with the
property that $n(\mu )\leq \mu^{-1}$ and, for all $q\in \Gamma^{\ast }$, $%
\min_{i=1,2,\ldots ,n(\mu )}{|q-q_{i}|}\leq \mathrm{const}\,\mu^{1/d}$. If
the subspace $S\subset \ell^{2}(\Gamma )$ has dimension larger than $%
(d+1)\mu^{-1}$ then there is a vector $\tilde{{\varphi }}\in S$ with $|%
\tilde{{\varphi }}|_{2}=1$ and 
\begin{equation}
\tilde{{\varphi }}\in \bigcap_{j=1}^{n(\mu )}\mathrm{Ker}\,\zeta
_{q_{j}}^{\prime }.
\end{equation}

By Taylor expansions, for such a vector $\tilde{{\varphi }}$ we have,
similarly as in the proof of (i), that for some constant $\mathrm{const}%
<\infty $ and all $p\in \Gamma^{\ast }$: 
\begin{eqnarray}
|\mathcal{F}^{\ast }\circ V^{1/2}\tilde{{\varphi }}(p)| &\leq &\mathrm{const}%
\,|V|_{1/2,2}^{1/2}\prod_{i=1}^{m}(1-\cos (p-p_{i})),  \label{b.2.Vphi} \\
|\mathcal{F}^{\ast }\circ V^{1/2}\tilde{{\varphi }}(p)| &\leq &\mathrm{%
const\,}\mu \,|V|_{1/2,2}^{1/2}.
\end{eqnarray}%
Using the last two inequalities we get 
\begin{eqnarray}
\lefteqn{|\langle \tilde{\varphi}\,|\,V^{1/2}h(\mathfrak{e})^{-1}V^{1/2}%
\tilde{\varphi}\rangle |}  \notag \\
&\leq &|\mathcal{F}^{\ast }\circ V^{1/2}\tilde{{\varphi }}|_{\infty
}\int_{\Gamma^{\ast }}\frac{|\mathcal{F}^{\ast }\circ V^{1/2}\tilde{{\varphi 
}}(p)|}{\mathfrak{e}(p)}\mathrm{d}\mu^{\ast }(p)  \notag \\
&\leq &\mathrm{const}\,\,\mu |V|_{1/2,2}.
\end{eqnarray}%
Thus, by (i), Lemma~\ref{lem-2.2} and the min-max principle, for some $%
\mathrm{const}<\infty $, $H(\mathfrak{e},V)$ has at most $(\mathrm{const}%
\,|V|_{1/2,2}+m)$ eigenvalues below $0$. \hfill $\square $

\begin{corollary}[{Semi-classical upper bound on $N{[e,V]}$ for $d=1,2$}]
\label{cor-5.5} Let $d\in \{1,2\}$ and $\mathfrak{e}$ be any admissible
dispersion relation from $C^{3}(\Gamma ^{\ast })$. Then there is a constant $%
c(\mathfrak{e})<\infty $ such that for all potentials $V\geq 0$, 
\begin{equation}
N[\mathfrak{e},V]\leq c(\mathfrak{e})(1+N_{sc}[\mathfrak{e},\tilde{V}]),
\label{semicld12}
\end{equation}%
where the effective potential $\tilde{V}$ is given by $\tilde{V}(x)\doteq
V(x)|x|^{d+5}$.
\end{corollary}

\noindent \textit{Proof:} From Lemma~\ref{lem-5.4} and Corollary~\ref%
{cor-3.2}: 
\begin{eqnarray}
N[\mathfrak{e},V] &\leq &|\{x\in \Gamma \,|\,\langle x\rangle^{d+5}V(x)\geq {%
\mathfrak{e}_{\mathrm{max}}}\}|+\#\mathrm{Min}(\mathfrak{e})  \notag \\
&&+C_{\ref{lem-5.4}(ii)}\left( \sum_{x\in \Gamma ,\,\langle x\rangle
^{d+5}V(x)<{\mathfrak{e}_{\mathrm{max}}}}\langle x\rangle^{-\frac{d+1}{2}%
}[\langle x\rangle^{d+1}\langle x\rangle^{4}V(x)]^{1/2}\right)^{2}.  \notag
\end{eqnarray}%
Thus, by the Cauchy-Schwarz inequality: 
\begin{eqnarray}
N[\mathfrak{e},V] &\leq &|\{x\in \Gamma \,|\,\langle x\rangle^{d+5}V(x)\geq {%
\mathfrak{e}_{\mathrm{max}}}\}|+\#\mathrm{Min}(\mathfrak{e})  \notag \\
&&+C_{\ref{lem-5.4}(ii)}\left( \sum_{x\in \Gamma }\langle x\rangle
^{-(d+1)}\right) \left( \sum_{x\in \Gamma ,\,\langle x\rangle^{d+5}V(x)<{%
\mathfrak{e}_{\mathrm{max}}}}\langle x\rangle^{d+5}V(x)\right) .  \notag
\end{eqnarray}%
As $\mathfrak{e}$ is a Morse function this implies \eqref{semicld12} in the
case $d=2$. Observing that $\langle x\rangle^{d+5}V(x)\leq \lbrack ${$%
\mathfrak{e}$}${_{\mathrm{max}}}\langle x\rangle^{d+5}V(x)]^{1/2}$, whenever 
$\langle x\rangle^{d+5}V(x)\leq ${$\mathfrak{e}$}${_{\mathrm{max}}} $, the
case $d=1$ follows from the last inequality as well. \hfill $\square $

\appendix

\section{Appendix \label{sec-A}}

\subsection{Proof of Lemma \protect\ref{lem-2.2} and Theorem \protect\ref%
{thm-2.6}\label{subsec-A.2}}

\noindent \textbf{Proof of Lemma~\ref{lem-2.2}:} We recall that, due to the
compactness of $V$, the Birman-Schwinger operator $B(\rho )$ is compact and
has only discrete spectrum above $0$. Similarly, the spectrum of $H(%
\mathfrak{e},V)$ below $0$ is discrete because $-V=H(\mathfrak{e},V)-H(%
\mathfrak{e},0)$ is compact.

Suppose that $-\rho <0$ is an eigenvalue of $H(\mathfrak{e},V)$ of
multiplicity $M\in \mathbb{N}$ and let $\{{\varphi }_{1},\ldots ,{\varphi }%
_{M}\}\subseteq \ell^{2}(\Gamma )$ be an ONB of the corresponding
eigenspace. Set 
\begin{equation}
\psi _{1}\doteq V^{1/2}{\varphi }_{1},\ldots ,\psi _{M}\doteq V^{1/2}{%
\varphi }_{M}.  \label{eq-2.4}
\end{equation}%
Then $\psi _{m}\in \ell^{2}(\Gamma )$ since $V\in \ell^{\infty }(\Gamma )$.
Moreover, 
\begin{equation}
{\varphi }_{m}\ =\ [\rho +h(\mathfrak{e})]^{-1}V{\varphi }_{m}\ =\ [\rho +h(%
\mathfrak{e})]^{-1}V^{1/2}\psi _{m},  \label{eq-2.5}
\end{equation}%
and the boundedness of $[\rho +h(\mathfrak{e})]^{-1}V^{1/2}$ implies that $%
\{\psi _{1},\ldots ,\psi _{M}\}\subseteq \ell^{2}(\Gamma )$ is linearly
independent. Clearly, \eqref{eq-2.4} and \eqref{eq-2.5} also yield 
\begin{equation}
B(\rho )\psi _{m}\ =\ V^{1/2}[\rho +h(\mathfrak{e})]^{-1}V^{1/2}\psi _{m}\
=\ \psi _{m},  \label{eq-2.6}
\end{equation}%
and hence the eigenspace of $B(\rho )$ corresponding to the eigenvalue $1$
has at least dimension $M$.

Conversely, if $\{\psi _{1},\ldots ,\psi _{L}\}\subseteq \ell^{2}(\Gamma )$
is an ONB of the eigenspace of $B(\rho )$ corresponding to the eigenvalue $1$
then we set 
\begin{equation}
{\varphi }_{1}\doteq \lbrack \rho +h(\mathfrak{e})]^{-1}V^{1/2}\psi
_{1},\ldots ,{\varphi }_{L}\doteq \lbrack \rho +h(\mathfrak{e}%
)]^{-1}V^{1/2}\psi _{L}.  \label{eq-2.7}
\end{equation}%
Since $[\rho +h(\mathfrak{e})]^{-1}V^{1/2}$ is bounded, ${\varphi }_{\ell
}\in \ell^{2}(\Gamma )$. Moreover, 
\begin{equation}
\psi _{\ell }\ =\ B(\rho )\psi _{\ell }\ =\ V^{1/2}{\varphi }_{\ell },
\label{eq-2.8}
\end{equation}%
and the boundedness of $V^{1/2}$ implies that $\{{\varphi }_{1},\ldots ,{%
\varphi }_{L}\}\subseteq \ell^{2}(\Gamma )$ is linearly independent.
Clearly, \eqref{eq-2.7} and \eqref{eq-2.8} also yield 
\begin{equation}
H(\mathfrak{e},V){\varphi }_{\ell }\ =\ -\rho {\varphi }_{\ell },
\label{eq-2.9}
\end{equation}%
and hence the eigenspace of $H(\mathfrak{e},V)$ corresponding to the
eigenvalue $-\rho $ has at least dimension $L$.

These arguments prove (i) and (ii) and, furthermore, $M=L$ and thus (iii),
i.e., 
\begin{equation}
\forall \, \rho >0: \ \dim \ker \big[H(\mathfrak{e},V)+\rho \big]\ =\ \dim
\ker \big[B(\rho )-1\big].  \label{eq-2.10}
\end{equation}

Observe that for all $\rho^{\prime }, \rho$ with $\rho^{\prime }\geq \rho >0$%
: $B(\rho^{\prime })\leq B(\rho )$. As the map $\rho \mapsto B(\rho )$ is
norm continuous on $%
\mathbb{R}
^{+}$ and $\lim\limits_{\rho \rightarrow \infty }B(\rho )=0$, by the min-max
principle, if $z_{k}>1$ is the $k$--th eigenvalue of $B(\rho )$ counting
from above with multiplicities, then there is a $\rho _{k}>\rho $ such that $%
1$ is the $k$--th eigenvalue of $B(\rho _{k})$ (counting from above with
multiplicities). Clearly, $\rho _{k^{\prime }}\leq \rho _{k}$, whenever $%
k^{\prime }\geq k$. By (iii), this implies that $H(\mathfrak{e},V)$ has at
least as many eigenvalues less or equal $-\rho $ as $B(\rho )$ has
eigenvalues greater or equal 1. By similar arguments, $B(\rho )$ has at
least as many eigenvalues greater or equal 1 as $H(\mathfrak{e},V)$ has
eigenvalues less or equal $-\rho $.\newline
{\phantom{A}}\hfill $\square $\bigskip

To prove Theorem \ref{thm-2.6}, we use the following estimate derived in 
\cite{Frank2014}:

\begin{proposition}[Frank]
\label{prop Frank}Let $(X,\mu )$ be any $\sigma $-finite measure space and $%
T $ a positive selfadjoint operator on $L^{2}(X,\mathbb{C})$ whose kernel is
trivial. Assume that there are given constants $\nu >2$ and $C_{\ref{prop
Frank}}\in \mathbb{R}^{+}$, such that, for all $E>0$ and any measurable set $%
\Omega \subset X$, 
\begin{equation*}
\mathrm{Tr}\left( \chi _{\Omega }T^{-1}\mathbf{1}[T\in (0,E]]\chi _{\Omega
}\right) \leq c_{\ref{prop Frank}}\mu (\Omega )E^{\frac{\nu -2}{2}},
\end{equation*}%
where $\chi _{\Omega }$ is the multiplication operator with the
characteristic function of $\Omega $. Let $V$ be any bounded positive-valued
measurable function and denote by $N(T,V)$ the number of discrete negative
eigenvalues, counting multiplicities, of the selfadjoint operator $T-V$. Then%
\begin{equation*}
N(T,V)\leq \frac{C_{\ref{prop Frank}}\nu }{2}\left( \frac{\nu }{\nu -2}%
\right) ^{\nu -2}\int_{X}V(x)^{\frac{\nu }{2}}\mathrm{d}\mu (x).
\end{equation*}
\end{proposition}

\noindent Observe that the above proposition is only a special case of \cite[%
Theorem 3.2]{Frank2014}.\medskip

\noindent \textbf{Proof of Theorem~\ref{thm-2.6} (CLR Bound):} Let $d\geq 3$
and take, in the above proposition, $X\doteq \mathbb{Z}^{d}$, $\mu $ as
being the counting measure, and $T\doteq h(\mathfrak{e})$. Then, clearly,%
\begin{equation*}
\mathrm{Tr}\left( \chi _{\Omega }T^{-1}\mathbf{1}[T\in (0,E]]\chi _{\Omega
}\right) \leq \mu (\Omega )\int_{\mathfrak{e}^{-1}((0,E])}\frac{1}{\mathfrak{%
e}(p)}\mathrm{d}\mu ^{\ast }(p)\text{ },
\end{equation*}%
where we recall that $\mu ^{\ast }$ is the (normalized) Haar measure on the $%
d$-dimensional torus $\Gamma ^{\ast }$. If the dispersion $\mathfrak{e}$ is
a Morse function then $\mathfrak{e}(p)=\mathcal{O}(|p-p_{0}|^{2})$ near
momenta $p_{0}\in \Gamma ^{\ast }$ minimizing\ $\mathfrak{e}$ and, hence, 
\begin{equation*}
\int_{\mathfrak{e}^{-1}((0,E])}\frac{1}{\mathfrak{e}(p)}\mathrm{d}\mu ^{\ast
}(p)\leq C_{\ref{prop Frank}}E^{\frac{d-2}{2}}
\end{equation*}%
for some $C_{\ref{prop Frank}}\in \mathbb{R}^{+}$ and all $E>0$. Observe
that this constant can be chosen uniformly w.r.t. $\Vert \mathfrak{e}\Vert
_{C^{3}}$ and $K(\mathfrak{e})$. The theorem directly follows from these two
estimates combined with Proposition \ref{prop Frank}.\hfill $\square
\medskip $

\subsection{Proof of Lemma \protect\ref{lem-3.1}, Theorem \protect\ref%
{thm-3.5} and Lemma \protect\ref{l.b.cont} \label{subsec-A.3}}

\textbf{\noindent Proof of Lemma~\ref{lem-3.1}:} We assume that $%
N[B],N[A]<\infty $, otherwise there is nothing to prove. As $N[A+B]\leq
N[A-B_{-}]$ and $N[B]=N[-B_{-}]$, it suffices to show that%
\begin{equation*}
N[A-B_{-}]\ \leq \ N[A]+N[-B_{-}].
\end{equation*}%
Here, $B_{-}\doteq |B|\,\mathbf{1}[B<0]$. Let $M\doteq N[B]=\mathrm{\dim
\;Ran}(B_{-})$ and assume that $A-B_{-}$ has at least $N[A]+M+1$ eigenvalues
(counting multiplicities) below $0$. Then, by the min-max principle, there
is a subspace $X\subset $ $\mathcal{H}$, $\mathrm{\dim }\;X=$ $N[A]+M+1$,
for which%
\begin{equation*}
\sup_{\psi \in X,\;|\psi |_{2}=1}\langle \psi \,|\,(A-B_{-})(\psi )\rangle
<0.
\end{equation*}%
Hence%
\begin{equation*}
\sup_{\psi \in X\cap \ker (B_{-}),\;|\psi |_{2}=1}\langle \psi
\,|\,(A-B_{-})(\psi )\rangle =\sup_{\psi \in X\cap \ker (B_{-}),\;|\psi
|_{2}=1}\langle \psi \,|\,A(\psi )\rangle <0.
\end{equation*}%
$\mathrm{\dim \;}X\cap \ker (B_{-})\geq \mathrm{\dim \;}X-M=$ $N[A]+1$.
Again by the min-max principle, this would then imply that $N[A]\geq $ $%
N[A]+1$. \hfill $\square $\bigskip

For any $\chi \in C^{\infty }({\mathbb{R}}^{d},{\mathbb{R}}),$ define its 
\textit{Gevrey norms} by: 
\begin{equation}
\Vert \chi \Vert _{s,R}\doteq \sum_{\underline{n}\in {\mathbb{N}}_{0}^{d}}%
\frac{R^{|\underline{n}|}}{(\underline{n}!)^{s}}\sup_{p\in {\mathbb{R}}%
^{d}}|\partial _{p}^{\underline{n}}\chi (p)|,\quad s\geq 1,\;R>0.
\label{Gev,norms}
\end{equation}%
The function $\chi $ is called \textit{$s$-Gevrey} if for some $R>0$, $\Vert
\chi \Vert _{s,R}<\infty $.

\begin{lemma}
\label{bound.TF.Gevrey} Let $\chi \in C_{0}^{\infty }({\mathbb{R}}^{d},{%
\mathbb{R}})$. Then, for all $p\in {\mathbb{R}}^{d}$, 
\begin{equation*}
|\hat{\chi}(p)|\leq \Vert \chi \Vert _{R,s}\,|\mathrm{supp\,}\chi |\,\exp
\left( 1-(\mathrm{e}^{-1}R|p|)^{\frac{1}{s}}\right) .
\end{equation*}%
Here, $|p|\doteq \max \{|p_{1}|,|p_{2}|,\ldots ,|p_{d}|\}$, $\hat{\chi}(p)
\doteq \int_{\mathbb{R}^d} e^{-ipx} \, \chi(x) \, \frac{d^dx}{(2\pi)^{d/2}}$
is the Fourier transform of $\chi$ on $\mathbb{R}^d$, and $|\mathrm{supp\,}%
\chi |$ is the volume of the support of the function $\chi $.
\end{lemma}

\noindent \textit{Proof}: The bound above is obvious if $\mathrm{e}%
^{-1}R|p|\leq 1$. Therefore, we only consider the case $\mathrm{e}%
^{-1}R|p|>1 $. By assumption, for all $n\in {\mathbb{N}}$: 
\begin{eqnarray}
|\hat{\chi}(p)| &\leq &\frac{(n!)^{s}}{(R\max \{|p_{1}|,|p_{2}|,\ldots
,|p_{d}|\})^{n}}\Vert \chi \Vert _{R,s}\,|\mathrm{supp\,}\chi |  \notag \\
&\leq &\frac{n^{sn}}{R^{n}|p|^{n}}\Vert \chi \Vert _{R,s}\,|\mathrm{supp\,}%
\chi |.
\end{eqnarray}%
Now use that for all $r$ with $\mathrm{e}^{-1}r>1$ 
\begin{equation*}
\min_{n\in {\mathbb{N}}} \bigg\{ \frac{n^{sn}}{r^{n}} \bigg\} \ \leq \
\max_{\xi \in \lbrack -1,0]+(\mathrm{e}^{-1}r)^{\frac{1}{s}}} \Big\{ \mathrm{%
e}^{\xi (s\log (\xi )-\log(r))} \Big\} \, .
\end{equation*}%
\hfill $\square $

\begin{lemma}[Poisson summation formula]
\label{lemma.Poisson} Let $\chi :{\mathbb{R}}^{d}\rightarrow {\mathbb{R}}$
be smooth and assume that $\mathrm{supp\,}\chi $ is compact. Define $\tilde{%
\chi}:\Gamma _{d}^{\ast }\rightarrow {\mathbb{C}}$ by 
\begin{equation*}
\tilde{\chi}([p])\doteq \sum_{x\in {\mathbb{Z}}^{d}}\chi (x)\mathrm{e}%
^{ip\cdot x}.
\end{equation*}
Then, for all $p\in \lbrack -\pi ,\pi )^{d}$, 
\begin{equation*}
\tilde{\chi}([p])=(2\pi )^{d/2}\sum_{q\in (2\pi {\mathbb{Z}})^{d}}\hat{\chi}%
(p+q).
\end{equation*}
\end{lemma}

\begin{corollary}
\label{corollary.Poisson} For all $p\in \lbrack -\pi ,\pi )^{d}$, all $R>1$,
and all $s \geq 1$, 
\begin{eqnarray}
\big| \tilde{\chi}([p])-(2\pi )^{\frac{d}{2}}\hat{\chi}(p) \big| & \leq &
(2\pi )^{\frac{d}{2}} \: \Vert \chi \Vert_{R,s} \: |\mathrm{supp\,}\chi | \:
\exp\big[ 1 - R^{\frac{1}{s}} \big] \: \sum_{p^{\prime }\in {\mathbb{Z}}%
^{d}} \exp\big[ \mathrm{e}^{-|p^{\prime }|^{\frac{1}{s}}} \big]  \notag \\%
[1ex]
& \leq & \mathrm{const}\,\Vert \chi \Vert _{R,s}\,|\mathrm{supp\,}\chi |\,\,
\exp\big[ - R^{\frac{1}{s}} \big] \: ,  \notag
\end{eqnarray}%
where $\mathrm{const}<\infty $ is a constant depending only on $s$ and $d$.
\end{corollary}

\noindent \textbf{Proof of Theorem~\ref{thm-3.5}:} For simplicity, we
temporarily assume that the hopping matrix $h(\mathfrak{e})$ has finite
range. Let $\chi \in C^{\infty }({\mathbb{R}},{\mathbb{R}})$ be any Gevrey
function with: $0\leq \chi (x)\leq 1$ for all $x\in {\mathbb{R}}$; $\chi
(x)=1$ for all $x$, $|x|\leq 1$; and $\chi (x)=0$ for all $x$, $|x|\geq 2$.
Such a $s$-Gevrey function exists for any $s>1$. For each $L,\Delta L>0$
define the Gevrey function $\tilde{\Phi}_{L,\Delta L}:{\mathbb{R}}%
^{d}\rightarrow {\mathbb{R}}$, 
\begin{equation}
\tilde{\Phi}_{L,\Delta L}(x)\ \doteq \ \chi \big((x_{1}+L)/\Delta L\big)\chi
(x_{2}/\Delta L)\cdots \chi (x_{d}/\Delta L).
\end{equation}%
If $\chi $ is a $s$-Gevrey function, by definition of the Gevrey norms, for
some $\mathrm{const}<\infty $, some $\Delta L_{0}>0$, and all $L,\Delta L>0$%
: 
\begin{equation}
\Vert \tilde{\Phi}_{L,\Delta L}\Vert _{s,\Delta L/\Delta L_{0}}\leq \mathrm{%
const}.  \label{bound.Gev.L}
\end{equation}

Let $p^{(0)}\in \mathrm{Min}(\mathfrak{e})$, i.e. $\mathfrak{e}(p^{(0)})=0$.
Define for each $L,\Delta L>0$, the vector $\Phi _{L,\Delta L}\in \ell
^{2}(\Gamma )$, 
\begin{equation}
\Phi _{L,\Delta L}(x)=\mathrm{e}^{ip_{0}\cdot x}\tilde{\Phi}_{L,\Delta
L}(x),\quad x\in \Gamma .
\end{equation}

By (\ref{bound.Gev.L}), Lemma~\ref{bound.TF.Gevrey} and Corollary \ref%
{corollary.Poisson}, for some constant $\mathrm{const}<\infty $ depending
only on $\mathfrak{e}$ and all $L,\Delta L\geq 1$: 
\begin{equation}
\left\vert \langle \Phi _{L,\Delta L}\,|\,h(\mathfrak{e})\Phi _{L,\Delta
L}\rangle \right\vert \leq \mathrm{const}\,\left( \Delta L\right)^{-2}|\Phi
_{L,\Delta L}|_{2}^{2}.  \label{b.L.e}
\end{equation}

Observe that, by the assumption (\ref{eq-3.16}), for some constant $\mathrm{%
const}>0$ and all $L,\Delta L\geq 1$: 
\begin{equation}
\langle \Phi _{L,\Delta L}\,|\,V\Phi _{L,\Delta L}\rangle \geq \mathrm{const}%
\,(L+\Delta L)^{-\alpha }|\Phi _{L,\Delta L}|_{2}^{2}.  \label{b.L.V}
\end{equation}%
Let $R<\infty $ be the range of the hopping matrix $h(\mathfrak{e})$. Notice
that, for all $L,\Delta L>0$ and all $L^{\prime },\Delta L^{\prime }>0$ with 
$L+2\Delta L+R<L^{\prime }-2\Delta L^{\prime }-R$, 
\begin{equation}
\langle \Phi _{L,\Delta L}\,|\,H(\mathfrak{e},V)\Phi _{L^{\prime },\Delta
L^{\prime }}\rangle =0.  \label{L.orth}
\end{equation}%
For any fixed $N\in {\mathbb{N}}$ and $L>0$, define $L_{k},\Delta L_{k}$, $%
k=1,2,\ldots ,N$, by: 
\begin{equation}
L_{k}=k\,L,\quad \Delta L_{k}=L/8.
\end{equation}%
Then, for $L$ sufficiently large, (\ref{L.orth}) is satisfied for all $%
(L,\Delta L)=(L_{k},\Delta _{k})$, $(L^{\prime },\Delta L^{\prime
})=(L_{l},\Delta _{l})$, $k\neq l$. Furthermore, by (\ref{b.L.e}) and (\ref%
{b.L.V}), as $\alpha <2$, for $L$ large enough: 
\begin{equation}
\langle \Phi _{L_{k},\Delta L_{k}}\,|\,H(\mathfrak{e},V)\Phi _{L_{k},\Delta
L_{k}}\rangle <0,\quad k=1,2,\ldots ,N.  \label{b123}
\end{equation}%
It follows by the min-max principle that for all $N\in {\mathbb{N}}$, $N[%
\mathfrak{e},V]\geq N$.

Now assume that $h(\mathfrak{e})$ is not necessarily finite range, but still
satisfies the bound in (\ref{eq-3.16}). Then, for some $\mathrm{const}%
<\infty $ not depending on $L$ and all $k,l=1,2\ldots ,N$, $k\neq l$, 
\begin{eqnarray}
\left\vert \langle \Phi _{L_{k},\Delta L_{k}}\,|\,H(\mathfrak{e},V)\Phi
_{L_{l},\Delta L_{l}}\rangle \right\vert &<&\mathrm{const}\,L^{-\alpha
^{\prime }}|\Phi _{L_{k},\Delta L_{k}}|_{2}|\Phi _{L_{l},\Delta L_{l}}|_{2} 
\notag \\
&=&\mathrm{const}\,L^{-\alpha^{\prime }}|\Phi _{L_{1},\Delta L_{1}}|_{2}^{2}.
\end{eqnarray}%
It follows from this bound, (\ref{b.L.e}), and (\ref{b.L.V}) that 
\begin{equation*}
\max\limits_{{\varphi }\in \mathrm{span}\{\Phi _{L_{1},\Delta L_{1}},\ldots
,\Phi _{L_{N},\Delta L_{N}}\},\;|{\varphi }|_{2}=1}\langle {\varphi }\,|\,H(%
\mathfrak{e},V){\varphi }\rangle \leq \mathrm{const}^{\prime }L^{-\alpha
^{\prime }}-\mathrm{const}\,L^{-\alpha }
\end{equation*}%
for some $\mathrm{const}>0$, $\mathrm{const}^{\prime }<\infty $ depending on 
$N$ but not on $L$. As, by assumption, $\alpha <\alpha^{\prime }$, the
right-hand side of the equation above is strictly negative for $L$
sufficiently large. Thus, by the min-max principle, for all $N\in {\mathbb{N}%
}$, $N[\mathfrak{e},V]\geq N$. \hfill $\square $\bigskip

\noindent \textbf{Proof of Lemma~\ref{l.b.cont}:} Let $\chi :{\mathbb{R}}%
\rightarrow {\mathbb{R}}_{0}^{+}$ be a smooth function with $\chi (x)=1$ if $%
|x-1/2|\leq 1/2$, and $\chi (x)=0$ if $|x-1/2|\geq 3/4$. We will assume that 
$\chi $ is a $s$--Gevrey function for some $s>1$. For all $M,m\in {\mathbb{N}%
}_{0}$, all $X\in {\mathbb{Z}}^{d}$, and all $\underline{k}\in \{0,1,\ldots
,2^{m}-1\}^{d}$ define the function $\Phi (M,m\,\,|\,X,\underline{k}):{%
\mathbb{R}}^{d}\rightarrow {\mathbb{R}}_{0}^{+}$ by 
\begin{equation}
\Phi (M,m\,|\,X,\underline{k})(y)\doteq \prod\limits_{i=1}^{d}\chi \left(
2^{M+m}(y_{i}-2^{-M}X_{i}-2^{-M-m}k_{i})\right) .
\end{equation}%
Clearly, if $(X,\underline{k})\not=(X^{\prime },\underline{k}^{\prime })$, 
\begin{equation}
\mathrm{dist}\left( \mathrm{supp}\,\Phi (M,m\,|\,X,\underline{k})\,,\mathrm{%
supp}\,\Phi (M,m\,|\,X^{\prime },\underline{k}^{\prime })\right) \geq
2^{-(M+m+2)}.
\end{equation}

Let $p^{(0)}\in \mathrm{Min}(\mathfrak{e})$ and let $c_{0}<\infty $ be some
constant such that for some $\epsilon >0$ and all $p\in B(p_{0},\epsilon )$, 
$\mathfrak{e}(p)\leq c_{0}|p-p^{(0)}|^{2}$. Let further $c_{1}$ be a
constant with 
\begin{equation}
\int_{{\mathbb{R}}^{d}}\,|p|^{2}\,|\hat{\Phi}(p)|^{2}\,\mathrm{d}^{d}p\leq
c_{1}\int_{{\mathbb{R}}^{d}}|\hat{\Phi}(p)|^{2}\,\mathrm{d}^{d}p,
\end{equation}%
where $\hat{\Phi}$ is the Fourier transform of $\Phi (0,0\,|\,0,0)$.

Let $\mathbf{X}\doteq \{X_{1},\ldots ,X_{N}\}$ be the set of points from ${%
\mathbb{Z}}^{d}$ on which 
\begin{equation}
2c_{0}c_{1}[2^{M+m_{n}}]^{2}<v_{-}^{(M)}(2^{-M}X_{n})\;\mathrm{for}\,\mathrm{%
some\;}m_{n}\geq 0.  \label{eq.mn}
\end{equation}%
For all $n\in \{1,\ldots ,N\}$ let $m_{n}\in {\mathbb{N}}_{0}$ be the
largest integer satisfying (\ref{eq.mn}).

For all $L>0$ define the functions $\Phi _{n,\underline{k}}^{(L)}\in \ell
^{2}(\Gamma )$, $n=\{1,2,\ldots ,N\}$, $\underline{k}\in \{0,1,\ldots
,2^{m_{n}}-1\}^{d}$ by 
\begin{equation}
\Phi _{n,\underline{k}}^{(L)}(x)\doteq \mathrm{e}^{ip_{0}\cdot x}\Phi
(M,m_{n}\,|\,X_{n},\underline{k})(L^{-1}x).
\end{equation}

Using Lemma~\ref{lemma.Poisson} we see that, by construction, for all $%
n=\{1,2,\ldots ,N\}$ and all $\underline{k}\in \{0,1,\ldots
,2^{m_{n}}-1\}^{d}$, 
\begin{eqnarray}
\lefteqn{\langle \Phi _{n,\underline{k}}^{(L)}\,|\,H(\mathfrak{e},V_{L})\Phi
_{n,\underline{k}}^{(L)}\rangle }  \notag \\
&\leq &\left[ -\frac{1}{2}L^{-2}v_{-}^{(M)}(2^{-M}X_{n})+\mathcal{O}(L^{-3})%
\right] |\Phi _{n,\underline{k}}^{(L)}|_{2}^{2}.
\end{eqnarray}%
Furthermore, for all $(n,\underline{k})$, $(n^{\prime },\underline{k}%
^{\prime })$, $n,n^{\prime }\in \{1,2,\ldots ,N\}$, $\underline{k}\in
\{0,1,\ldots ,2^{m_{n}}-1\}^{d}$, $\underline{k}^{\prime }\in \{0,1,\ldots
,2^{m_{n}^{\prime }}-1\}^{d}$ with $(n,\underline{k})\not=(n^{\prime },%
\underline{k}^{\prime })$, we have, for some $\mathrm{const}<\infty $ not
depending on $L$, the following estimate: 
\begin{equation}
|\langle \Phi _{n,\underline{k}}^{(L)}\,|\,H(\mathfrak{e},V_{L})\Phi
_{n^{\prime },\underline{k}^{\prime }}^{(L)}\rangle |\leq \mathrm{const}%
\,L^{-\alpha }|\Phi _{n,\underline{k}}^{(L)}|_{2}|\Phi _{n^{\prime },%
\underline{k}^{\prime }}^{(L)}|_{2}.
\end{equation}%
Finally, (\ref{lim.inf.continuum}) follows by using the min-max principle
and observing that, by the choice of the numbers $m_{n}$, for some $\mathrm{%
const}^{\prime }>0,$ 
\begin{equation*}
2^{dM}2^{dm_{n}}\geq \mathrm{const}^{\prime
}[v_{-}^{(M)}(2^{-M}X_{n})]^{d/2}.
\end{equation*}%
\hfill\ \hfill\ $\ \hfill \square $

\bigskip \noindent \textbf{Acknowledgements: }This research is supported by
the FAPESP (grant 2016/02503-8), the CNPq and the Spanish Ministry of Economy and
Competitiveness MINECO: BCAM Severo Ochoa accreditation SEV-2013-0323 and
MTM2014-53850. 


\end{document}